\documentclass[prl,twocolumn,showpacs,superscriptaddress]{revtex4-2} 
\usepackage{xcolor}
\usepackage{amsmath}
\usepackage{amssymb}   
\usepackage{amsthm}   
\usepackage{bm}
\usepackage{float}
\usepackage{url}
\usepackage{graphicx}
\usepackage{algorithm,algcompatible}
\usepackage{hyperref}
\hypersetup{
  colorlinks   = true, %Colours links instead of ugly boxes
  urlcolor     = blue, %Colour for external hyperlinks
  linkcolor    = black, %Colour of internal links
  citecolor   = black %Colour of citations
}

\newcommand{\jd}[1]{\textcolor{black}{#1}}

\begin{document}

\title{Topological packing statistics distinguish living and non-living matter}

\author{Dominic J. Skinner}
\affiliation{Department of Mathematics, Massachusetts Institute of Technology, 77 Massachusetts Avenue, Cambridge, MA 01239, USA}
\affiliation{NSF-Simons Center for Quantitative Biology, Northwestern University,\\2205 Tech Drive, Evanston, IL 60208, USA}
\author{Hannah Jeckel}
\affiliation{Department of Physics, Philipps-Universit\"at Marburg, Renthof 6, 35032 Marburg, Germany}
\affiliation{Biozentrum, University of Basel, Spitalstrasse 41, 4056 Basel, Switzerland}
\author{Adam C. Martin}
\affiliation{Department of Biology, Massachusetts Institute of Technology, 77 Massachusetts
Ave., Cambridge, MA 02139, USA}
\author{Knut Drescher}
\affiliation{Department of Physics, Philipps-Universit\"at Marburg, Renthof 6, 35032 Marburg, Germany}
\affiliation{Biozentrum, University of Basel, Spitalstrasse 41, 4056 Basel, Switzerland}
\author{J\"orn Dunkel}
\thanks{To whom correspondence should be addressed.\\ E-mail: dunkel@mit.edu}
\affiliation{Department of Mathematics, Massachusetts Institute of Technology, 77 Massachusetts Avenue, Cambridge, MA 01239, USA}

\begin{abstract}

How much structural information is needed to distinguish living from non-living systems? Here, we show that the statistical properties of Delaunay tessellations suffice to differentiate prokaryotic and eukaroytic cell packings from a wide variety of inanimate physical structures. By introducing a mathematical framework for measuring topological distances between general 3D point clouds, we construct a universal topological atlas encompassing  bacterial biofilms, \jd{snowflake yeast}, \jd{plant shoots}, zebrafish brain matter, \jd{organoids}, and embryonic tissues as well as \jd{foams}, colloidal packings, glassy materials, and stellar configurations. Living systems are found to localize within a bounded island-like region, reflecting that growth memory essentially distinguishes multicellular from physical packings. By detecting subtle topological differences, the underlying metric framework enables a unifying classification of  3D disordered media, from microbial populations,  organoids and tissues to amorphous materials and astrophysical systems.

\end{abstract}

\maketitle

Topology~\cite{carlsson2009topology} studies the fundamental  neighborhood relations amongst living~\cite{BialekFlocking, Hannezo2021, Couzin2011}, non-living~\cite{MacPhersonFoam,LazarPRL} or abstract~\cite{Barabasi2008,Strogatz2001} entities. By ignoring object-specific  features, such as particle shape or chemical composition, topological analysis of spatial packings and nearest-neighbor networks can reveal universal ordering principles~\cite{LazarRycroft, Goldstein2022} that extend across broad classes of systems~\cite{Stillinger84}.  As recent advances in high-resolution imaging~\cite{Keller2010, Cheng2019}, deep-learning based image analysis~\cite{Mu2021},  and simulation~\cite{FLETCHER2014,Lardon2011, 2018Sharp,Yan2022} techniques  are offering unprecedented insights into the spatial organization of biological~\cite{Keller2008, Keller2018,2021BiGuo_PNAS} and physical~\cite{DonevMandM} matter, there now exists a unique opportunity to explore the  topological similarities and differences across a diverse range of complex systems, from bacterial communities and eukaryotic tissues to amorphous materials~\cite{GoogleGlassy} or large-scale astrophysical structures~\cite{Joss2019}. A central open question in this context is whether, or to which extent, living and non-living matter have distinct topological properties. 

\par

To address this problem, we introduce here a general mathematical framework for comparing the topology of  three-dimensional (3D) disordered packings~\cite{LazarPRL} or, more generally, point clouds. Such point clouds can represent the positions of the cell nuclei in a piece of tissue~\cite{Keller2010, Sherf2017,2018Idema_JTB}, the midpoints of bacteria within a biofilm~\cite{Hartmann2019, Qin2020}, the atoms in a liquid or solid~\cite{Lazar2015, GoogleGlassy}, or the loci of nearby stars in our galaxy~\cite{1991GLIESE}. Extending  recent progress in the statistical characterization of topological structures~\cite{Skinner2021, LazarPRL, Lazar2015}, we developed a computationally  efficient algorithm that makes it possible to directly compare all the these and many other systems, \jd{without requiring curated training data}. Intuitively, the underlying numerical scheme generalizes the classical earth mover's distance~\cite{villani2009optimal}: In the first step,  the algorithm  determines the relative frequencies of typical  neighborhood patterns (\lq motifs\rq) within a given 3D point cloud; in the second step, it computes the cost of transforming one motif frequency distribution into another by exploiting a natural graph structure on the space of motifs. By applying this framework to a diverse set of  experimental and simulated data, we find that basic topological information suffices to identify  biofilms from  different bacterial  species (Fig.~1), to determine time-ordering and developmental transitions in  zebrafish embryos (Fig.~2), and to distinguish these and other living systems from a variety of inanimate physical structures (Fig.~3).  More generally, the statistical analysis approach developed here opens a path towards quantifying and comparing the structural differences within and across broad classes 3D disordered media.

\par

%%%%%%%%%%%%%%%%%%%
\begin{figure*}[t]%[\sidecaptionrelwidth][t]
\centering
\includegraphics{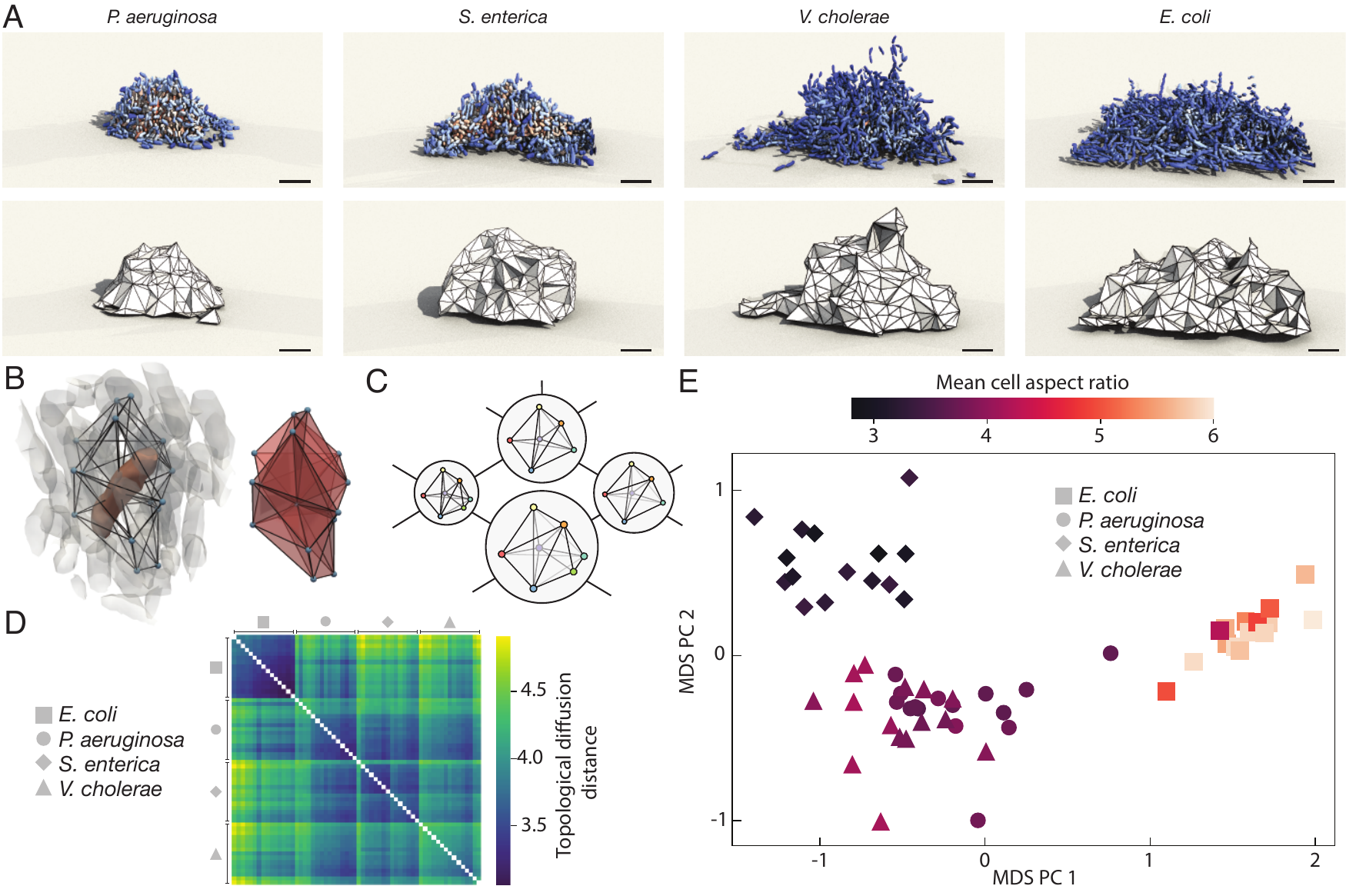}
\caption{\textbf{Topological diffusion distance (TDD) distinguishes prokaryotic multicellular colonies from local topological information alone.} 
(A)~3D reconstruction of bacterial biofilms formed by the species \emph{P. aeruginosa}, \emph{S. enterica}, \emph{V. cholerae} and \emph{E. coli} (top row; $\sim 2000$ cells per biofilm; experimental data from Ref.~\cite{BiofilmStructure}). The topological structure is encoded in (the $\alpha$-complex of) the Delaunay  tessellation (bottom row; SI Sec.~I). Color indicates local density (red:high, blue:low). Scale bar, 5$\,\mu$m. 
(B)~For each cell (red, left), we determine its nearest-neighbor motif (right), formed by all Delaunay tetrahedrons that have the cell and its neighbors as vertices (centroids of neighboring cells shown as blue spheres).
(C)~Motifs only change through discrete topological transitions (\lq flips\rq), which naturally induces a graph structure where each vertex is a motif and vertices are connected if they are one transition apart, illustrated here for selected motifs. Vertex size reflects the relative frequency of a motif. Each biofilm is thus mapped to a probability distribution of motifs over this flip-graph. 
(D)~Pairwise TDD matrix between all $4\times 15$ experiments, grouped by species. The block structure shows that the TDD detects differences between the species. 
(E)~Topological atlas obtained from the planar MDS embedding of the TDD distance matrix. The embedding is colored by the mean cell aspect ratio, showing that topological changes correlate strongly with changes in the cell geometry. TDD identifies every pair of biofilm species as statistically different at $p<0.01$ (SI Fig. S9).}
\label{fig:Top}
\end{figure*}
%%%%%%%%%%%%%%%%%%%

%%%%%%%%%%%%%%%%%%%
\begin{figure*}[t]%[\sidecaptionrelwidth][t]
\centering
\includegraphics{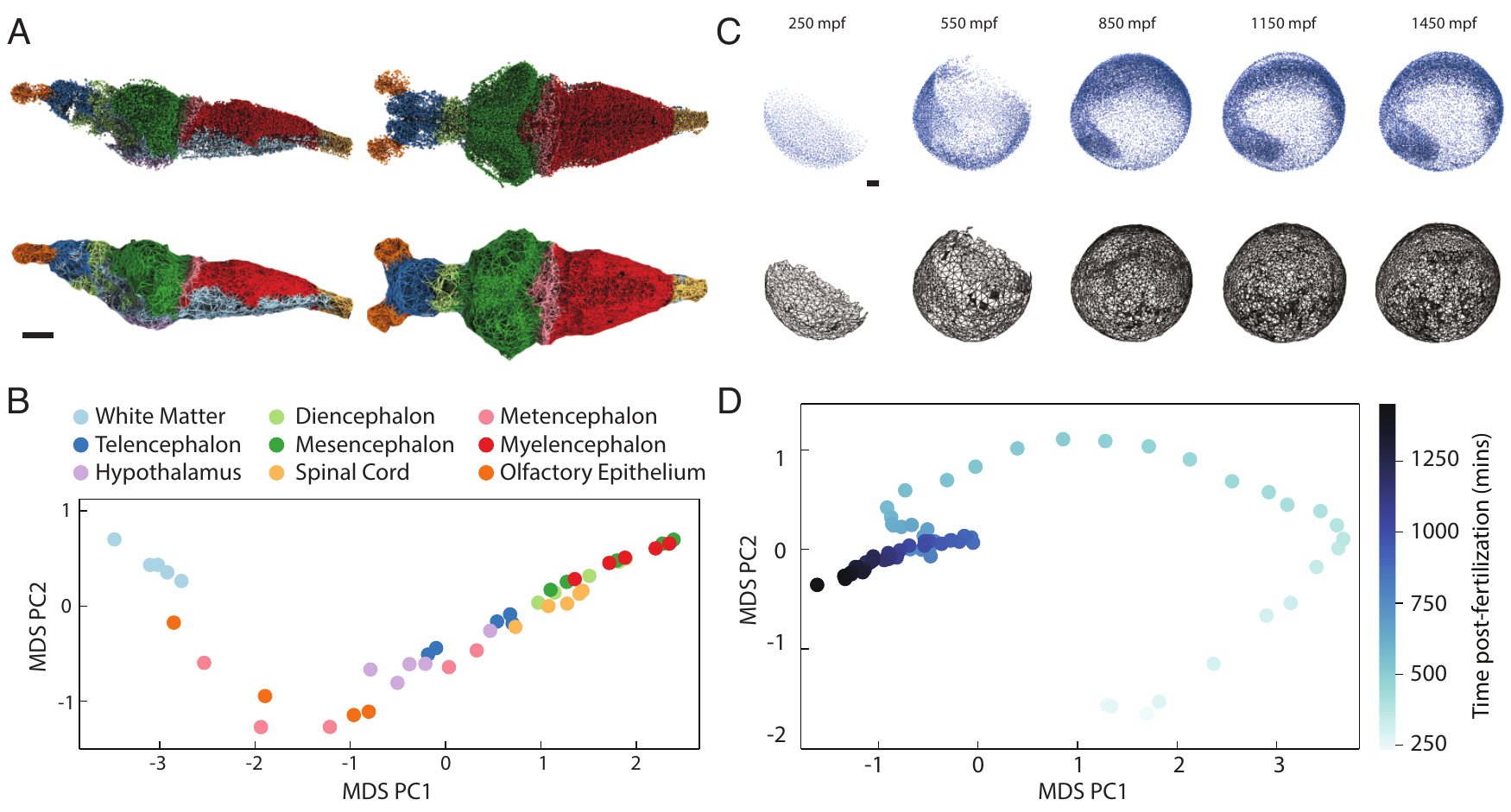}
\caption{\textbf{Topological statistics distinguish  static and dynamic eukaryotic architectures.} 
(A)~Nuclei within a juvenile zebrafish brain, as measured with X-ray micro-tomography, colored by 9 major brain regions (top), with corresponding Delaunay tessellation (bottom). Data is from Ref.~\cite{Cheng2019}, scale bar is 100$\mu m$. 
(B)~Topological analysis detects systematic differences between regions of the zebrafish brain (see also SI Fig.~S10), revealing that brain tissue architectures vary along a 1D topological manifold. The topological distance was computed pairwise between 9 brain regions across 5 separate experiments, and the resulting distance matrix was embedded with MDS. Although overall brain size and morphology differs across the experiments, corresponding regions of the brain  lie together in the embedding space. 
(C)~Cell nuclei in a zebrafish embryo imaged during development with light sheet microscopy (top), and the corresponding Delaunay tessellation (bottom). Data is from Ref.~\cite{Keller2008}, scale bar is 200$\mu$m (mpf: minutes post fertilization). 
(D)~Collecting the average topological distribution at 90 different time points and computing the pairwise distance matrix, the resulting MDS embedding recovers a curve parameterized by time. Note that, although the distance calculation and embedding are not explicitly aware of the temporal ordering of the data, the developmental progression is imprinted in the tissue topology and can thus be recovered from the TDD matrix.}
\label{fig:Fish}
\end{figure*}
%%%%%%%%%%%%%%%%%%%%%%%

Our starting point for constructing a metric framework that can measure distances between 3D disordered  structures is the classical Delaunay tessellation~\cite{LazarRycroft}, which is a topological object linking nearest neighbors in a point cloud.  
In three spatial dimensions, Delaunay tessellations are composed of adjacent  tetrahedrons,  illustrated in Fig.~1A for bacterial biofilm imaging data from four different species~\cite{BiofilmStructure}. If we pick any cell within a biofilm, then its nearest neighbors form an elementary  motif consisting of the corresponding Delaunay  tetrahedrons  (Fig.~1B). In a disordered multicellular system, these nearest-neighbor motifs typically  differ from cell to cell, but, as we show in SI Sec. II and SI Fig. S3, it is possible to uniquely identify and label each possible motif. This important fact allows us to count  how often each motif appears within a given 3D structure. The second and equally important mathematical observation is that one can determine how many elementary  neighbor  exchanges (\lq flips\rq) are needed to transform  one motif into another   (SI Sec. II). This means that we can construct a so-called \lq flip graph\rq{} ~\cite{Skinner2021} where each vertex represents a  specific motif and edges link motifs that are exactly one flip apart from each other (Fig.~1C). 
Any discrete 3D material structure can then be identified with a specific empirical probability distribution on the flip graph, by assigning to every motif its relative frequency in the material, indicated by the relative vertex size in Fig.~1C. In practice, a few hundred cells or particles often suffice to obtain a   sufficiently accurate approximation of the empirical motif probability distributions on the flip graph. 

\par
With the weighted flip graph representation at hand, measuring the topological distance between two disordered structures reduces to comparing their associated probability distributions on the flip graph. To do so, we implement a topological  diffusion distance (TDD) which approximates the well known earth mover's distance~\cite{Skinner2021}, or optimal transport, on a graph (SI Sec.~III), and use the TDD to compare structures of biological and non-biological matter. Intuitively, such transport distances measure how probability density needs to be shuffled along the flip graph to transform one motif distribution into  another. By utilizing information about the separation of motifs on the flip graph, such transport  distances are better at distinguishing disordered structures than conventional entropic distance measures  (SI Sec. III). The TDD has the practical advantage over other distances, that it can be efficiently computed, enabling a fast comparison of 3D structures even with several millions points and thousands of motifs. 

\par
As a first application, we find that the TDD is able to detect the subtle topological differences between biofilms formed by the bacterial species \emph{Pseudomonas aeruginosa}, \emph{Salmonella enterica}, \emph{Vibrio cholerae} and \emph{Escherichia coli} (Fig.~1D,E; 15 biofilm colonies per species with $\sim 2000$ cells each, data from Ref.~\cite{BiofilmStructure}). We used the 3D spatial position of cell centroids to compute a pairwise TDD distance matrix between colonies.  The block structure of the distance matrix  (Fig.~1D), and its 2D embedding  obtained using multi-dimensional scaling (MDS)~\cite{MDS} (Fig.~1E), show that the topological information encoded in the Delaunay tessellations suffices to distinguish the 3D structure of these four prokaryotic systems. In particular, we observe a close correlation between the mean cell aspect ratio and the topological clustering in the MDS atlas, which   illustrates that the TDD can  detect fine differences in positional and  orientational ordering arising from the interplay of steric repulsion and growth memory in bacterial biofilms~\cite{Hartmann2019,Pearce2019,Qin2020,BiofilmStructure}.

\begin{figure*}[t!]%[\sidecaptionrelwidth][t]
\centering
\includegraphics[width=\textwidth]{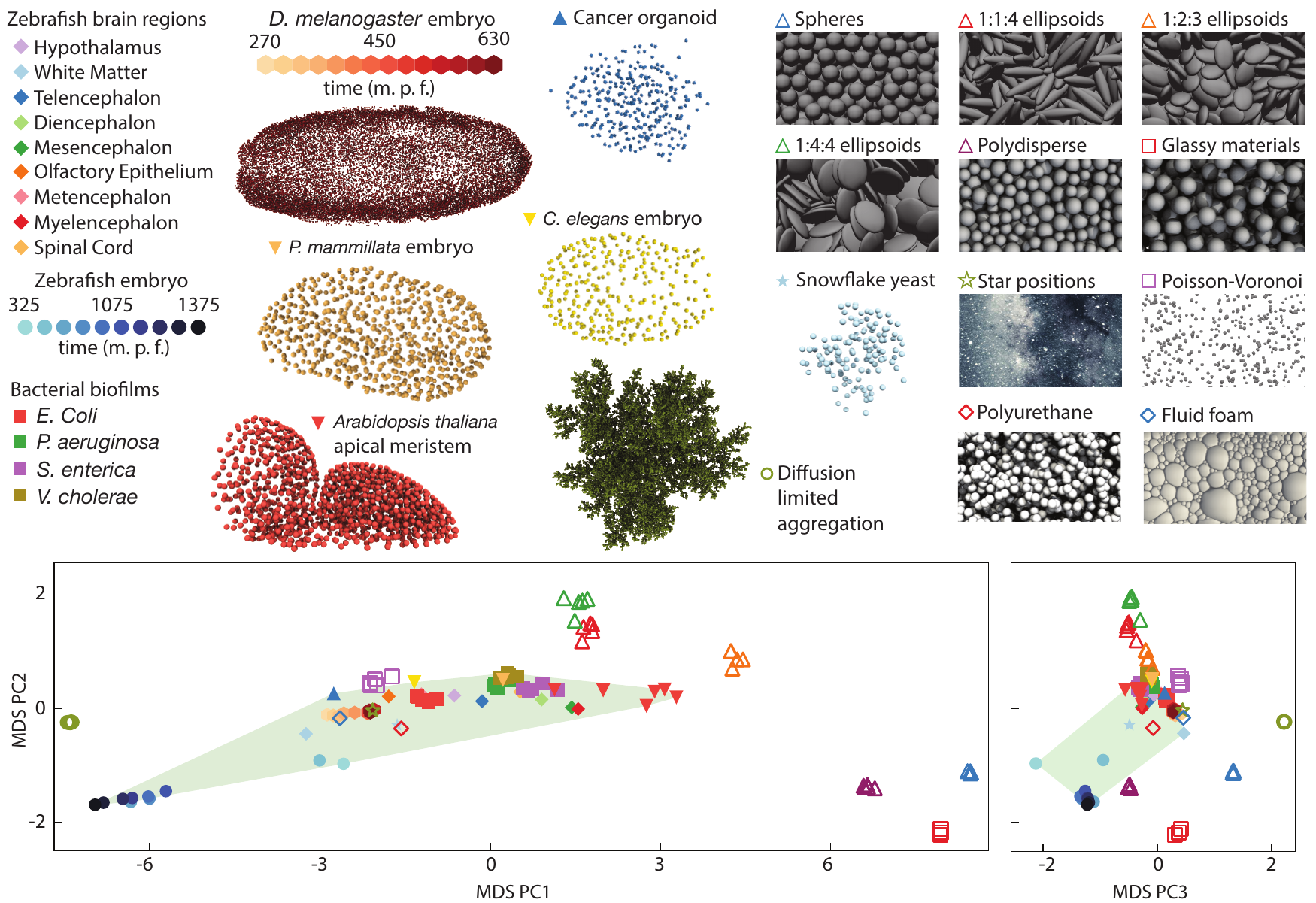}
\caption{\textbf{Combined topological atlas \jd{reveals topological variation across} living and non-living systems.} The atlas  was constructed by calculating the TDD between various biological and physical systems: bacterial biofilms, zebrafish brain regions, zebrafish embryo, fly   embryo (\emph{Drosophila melanogaster}), worm (\emph{Caenorhabditis elegans}) embryo,  sea squirt  embryo (\emph{Phallusia mammillata}), human cancer organoid, \jd{plant (\emph{Arabidopsis thaliana}) apical meristem, snowflake yeast, simulated random packings, a Polyurethane industrial foam, a simulated fluid foam}, a diffusion limited aggregation model,  Poisson-Voronoi tessellations, and a collection of stars close to the Earth (SI Table S1). The first three principal components of the MDS  embedding are shown, using filled symbols for living systems and non-filled symbols for non-living systems. For the various data sets  analyzed here, the convex hull enclosing the first three MDS components of the living systems (green) contains \jd{the industrial foam as the only non-living systems (1 out of 40 non-living points)}. 
}
\label{fig:PhaseDiag}
\end{figure*}

To demonstrate  the practical potential of the TDD framework for characterizing both static and dynamic tissue architecture in more complex eukarotic organisms, we next analyze the topology of brain and embryo tissues. State-of-the-art microscopy and image analysis methods are able to visualize  and segment the cell nuclei within the brains of commonly studied model organisms~\cite{Cheng2019,Mu2021}.  An open question is whether spatially and functionally distinct brain regions possess different topological characteristics and, if so,  what the properties of the underlying topological  manifold are.  As a specific example, we consider the juvenile zebrafish brain, for which $\sim 80,000$  nuclei positions were recently measured~\cite{Cheng2019} with X-ray micro-tomography (Fig.~\ref{fig:Fish}A). Computing the TDD matrix (SI Fig. 12), and its 2D MDS embedding, for 9 different brain regions across 5 experiments~\cite{Cheng2019}  reveals substantial topological differences between the white matter, hypothalamus and diencephalon regions (Fig.~\ref{fig:Fish}B). Interestingly, however, all 9 brain regions  localize near a 1D manifold (Fig.~\ref{fig:Fish}B), suggesting that topological variability in the juvenile  zebrafish brain tissue is highly constrained and can be effectively described by a single differentiation parameter along this manifold~\cite{Alba2021}.

\par
In addition to comparing static multicellular structures, the TDD makes it possible to quantify the topological changes of tissues during embryonic development. Recent advances in light-sheet microscopy enable the non-destructive imaging of tissue dynamics at a fine temporal resolution~\cite{Keller2008, Keller2010}. An open  question is whether topological information suffices to detect and define structural transitions during embryogenesis. Taking zebrafish development as a widely studied example, we analyzed lightsheet microscopy data from Ref.~\cite{Keller2008}, which reported the positions of all cell nuclei from around 200 to 1,500 minutes post fertilization (mpf), during which the number of cells increases from a few hundreds to around 15,000. As cells divide and rearrange to form different tissue domains, the internal structure of the Delaunay tessellations changes in time.  Five snapshots showing the evolution of the Delaunay networks of a zebrafish embryo are shown in Fig.~\ref{fig:Fish}C; in total, 900 time points were imaged at regular intervals separated by 90s. Dividing the data into 90 time intervals, each containing 10 sequential time points, we computed the $90\times 90$ TDD matrix (SI Fig. S13). The planar MDS embedding shows the topological trajectory of the zebrafish embryo (Fig.~\ref{fig:Fish}D). Note that, even though the TDD does not explicitly use temporal information,  one can recover the temporal ordering of the imaging data by following the topological trajectory reconstructed from the TDD. In other words, non-equilibrium growth memory is ingrained into the Delaunay structure. Therefore, changes in the topological trajectory can be used to define an intrinsic topological clock of the embryo, reminiscent of the proper-times derived from the particle worldlines in special and general relativity~\cite{Gravitation}. 

\par
An interesting mathematical aspect of the topological distance   framework, with substantial future theoretical and practical potential, is that the TDD  provides a foundation for developing a comprehensive geometric characterization of topological trajectories (as in  Figs.~\ref{fig:Fish}B,D) as well as  higher-dimensional embedding manifolds (such as in Fig.~\ref{fig:Top}E). This can be achieved combining the TDD approach with known  ideas and results from distance geometry~\cite{Liberti2016}, a branch of mathematics initiated by Hero of Alexandria about 2000 years ago  and advanced by Arthur Cayley and Karl Menger over the last two centuries
(SI Sec. IV, SI Fig. S8). Intuitively, distance geometry makes it possible to introduce well-defined notions of curvature and other geometric concepts (SI Sec. IV) on abstract point sets endowed with a metric structure; for instance, in Fig.~\ref{fig:Fish}D,  each point represents an ensemble of  Delaunay tessellations (developmental states of the zebrafish embryo) and a metric is provided by the TDD. By measuring the distances between three nearby points along the topological trajectory in Fig.~\ref{fig:Fish}D, one can compute the local Menger curvature of the developmental trajectory, corresponding to the inverse radius of the circumcircle of the three points (SI Fig.~S6). An application to zebrafish data~\cite{Keller2008} suggests that regions of extremal topological curvature correlate with structural transitions during biological development (SI Sec.~IV). This example illustrates how the TDD approach introduced  here can open new avenues for analyzing and comparing 3D disordered systems within and across disciplines.
\par
Indeed, arguably one of the most interesting applications of the TDD framework concerns our initial question:  Are the topological  architectures of living systems \jd{typically} distinct from those of non-living systems? Since the TDD does not require system-specific information beyond 3D Delaunay tessellations, it can be used to compute the topological distance between any pair of systems for which such tessellations are available. To initiate a   cross-disciplinary comparison, we determined pairwise TDDs between various biological and physical systems (SI Sec.~VI and SI Table S1), including bacterial biofilms~\cite{BiofilmStructure},  embryonic tissues from zebrafish~\cite{Keller2008}, worms~\cite{Cao2020}, sea squirts~\cite{AscidianScience}, flies~\cite{Keller2010}, \jd{human cancer organoids~\cite{GuoOrganoid}}, \jd{snowflake yeast~\cite{Goldstein2022}, plant shoot tips~\cite{Jonsson2016}}, random Poisson-Voronoi point sets, diffusion-limited aggregation structures, simulated granular packings~\cite{DonevMandM}, \jd{fluid and industrial foams~\cite{OpenCellFoam, FoamSimulation}}, and stars in our galaxy~\cite{1991GLIESE}. Remarkably, the resulting combined topological atlas (3D MDS embedding) in Fig.~3 suggests that the topological architectures of   biological systems are \jd{typically distinct from those of many  non-living physical systems, with the exception of industrial foams}.

\par

In conclusion, our above analysis shows that the statistical properties of Delaunay tessellations suffice to distinguish representative  prokaryotic and eukaroytic multicellular packings  from a wide variety of ordered and disordered physical structures. Structural memory, arising from positional and orientational correlations inherited during cell division and growth, apparently leads to characteristically different and  temporally evolving neighborhood motif distribution in biological systems. This hypothesis is supported by recent live-imaging experiments that demonstrated the importance of growth-induced ordering in bacterial  ~\cite{Hartmann2019,Pearce2019,Qin2020,BiofilmStructure, Volfson2008}  and  eukaryotic systems~\cite{Cohen2022}.
In the future, the topological distance framework introduced here can help overcome major current challenges in the analysis of natural and engineered multicellular structures, from quantifying how genetic mutations, diseases and drugs modify tissue architectures~\cite{1998Crompton_Review} to the  evaluation and classification of tumors~\cite{2021Wortman_npj} and  organoids~\cite{GuoOrganoid,2018Rossi_Nature}. Unlike machine learning approaches, which are  biased by the choice of training data and may fail to generalize~\cite{MEHTA2019}, the topological approach requires no training data, only the data to be compared. As more and more high-resolution 3D imaging data becomes available in the next years, they can be added to the topological atlas (Fig.~3) through the online platform~\cite{online_platform} provided with this work.  A particularly interesting challenge will be the identification and characterization of further exceptional biological or physical  systems that cross the topological boundaries between living and non-living matter.

\section{Acknowledgements}

\begin{acknowledgements}
This research was supported by a MathWorks Science Fellowship (D.J.S), NSF Award DMS-1764421 (D.J.S.), Simons Foundation grant 597491 (D.J.S.), the MIT Mathematics Robert E. Collins Distinguished Scholar Fund (J.D.), Sloan Foundation grant G-2021-16758 (J.D.),  MIT John W. Jarve (1978) Seed Fund for Science Innovation (A.C.M. and J.D.), the European Research Council (StG-716734), Deutsche Forschungsgemeinschaft DR 982/5-1 and DR 982/6-1 (K.D.), Minna James Heineman Foundation, Bundesministerium für Bildung und Forschung  TARGET-Biofilms (K.D.), the Swiss National Science Foundation NCCR “AntiResist” grant 51NF40$\_$180541 (K.D.), and fellowships from the Studienstiftung des deutschen Volkes and Joachim Herz Foundation (H.J.). We thank Ming Guo and Wenhui Tang for sharing organoid data from Ref.~\cite{GuoOrganoid}, \jd{Kerstin Weinberg and Stefan Buchen for sharing polyurethane data from Ref.~\cite{OpenCellFoam}, Thomas Day, Peter Yunker, and Raymond Goldstein for sharing snowflake yeast data from Ref.~\cite{Goldstein2022}, Henrik J\"onsson for sharing data from Ref.~~\cite{Jonsson2016}}, and
Justin Solomon for helpful discussions.
\end{acknowledgements}

\section{Author contributions}
D.J.S and J.D. developed the theoretical concepts and wrote the manuscript. D.J.S. performed all the analytical and numerical calculations. H.J., A.C.M. and K.D. provided data and input on the manuscript.

\section{Competing interests}
The authors declare no competing interests.

% Bibliography

\end{document}

\title{Supplementary information:\\ Topological packing statistics distinguish living and non-living matter}

\author{Dominic J. Skinner}
\affiliation{Department of Mathematics, Massachusetts Institute of Technology, 77 Massachusetts Avenue, Cambridge, MA 01239, USA}
\affiliation{NSF-Simons Center for Quantitative Biology, Northwestern University,\\2205 Tech Drive, Evanston, IL 60208, USA}
\author{Hannah Jeckel}
\affiliation{Department of Physics, Philipps-Universit\"at Marburg, Renthof 6, 35032 Marburg, Germany}
\affiliation{Biozentrum, University of Basel, Spitalstrasse 41, 4056 Basel, Switzerland}
\author{Adam C. Martin}
\affiliation{Department of Biology, Massachusetts Institute of Technology, 77 Massachusetts
Ave., Cambridge, MA 02139, USA}
\author{Knut Drescher}
\affiliation{Department of Physics, Philipps-Universit\"at Marburg, Renthof 6, 35032 Marburg, Germany}
\affiliation{Biozentrum, University of Basel, Spitalstrasse 41, 4056 Basel, Switzerland}
\author{J\"orn Dunkel}
\thanks{To whom correspondence should be addressed.\\ E-mail: dunkel@mit.edu}
\affiliation{Department of Mathematics, Massachusetts Institute of Technology, 77 Massachusetts Avenue, Cambridge, MA 01239, USA}

\maketitle

\tableofcontents

\section{Topological description}
\subsection{Delaunay tessellation}
Our starting point is a set of points, $\{x_i\}$, with $x_i \in \mathbb{R}^3$, with which we want to extract topological information from. Here, we do so by using the Delaunay tessellation, the dual of the Voronoi diagram. Recall that the Voronoi diagram divides space into regions associated with each point, $x_i$, so that the $i^{th}$ region is the set $V_i = \{ y\,  |\ ||y -x_i||^2_2 < ||y -x_j||_2^2, \forall j\neq i \}$, a definition which can be readily extended for $x_i\in\mathbb{R}^n$. Two points are connected in the Delaunay tessellation if their regions of the Voronoi diagram share a face. Moreover the Delaunay tessellation defines a simplicial complex, and is completely specified by a set of tetrahedrons in 3D (or triangles in 2D)~\cite{VoronoiBook}. This simplicial complex is the central object which we will extract information about our system from. 

\subsection{Local motifs}
Even for two experiment performed under exactly the same conditions, we would not expect the exact same realization of the Delauany tessellation. Instead we seek to characterize the statistical properties of the topological object, and we do so by quantifying local structure. Specifically, we define the local, or egocentric, simplex of radius $r$ around a point as the simplicial complex induced by all simplices consisting of points at most $r$ edges away from the vertex. We also refer to this as a motif, and consider a motif to describe the local neighborhood topology around a point. Every material could then be considered as a probability distribution over the space of motifs, where characterizing the differences between distributions characterizes the topological difference between materials.

The choice of $r$ determines how much information about the neighborhood structure is recorded at each point, but also how many observations are needed to sample the distribution well, and how computationally expensive the numerical problem of calculating the flip graph and distances will be. In 2D, only $O(20)$ distinct motifs of radius $r=1$ were typically observed and the flip graph was essentially a line graph~\cite{Skinner2021}. In this case $r\geq 2$ was required, and $r=2$ worked in practice, with some $O(50,000)$ unique motifs observed for $O(700,000)$ points in total~\cite{Skinner2021}. In 3D, even with $r=1$, $O(300,000)$ unique motifs were observed for some $O(700,000)$ points, and the flip graph is non-trivial. This suggests that taking $r=1$ is sufficient for 3D calculations. It also suggests that we are undersampling the true distribution. Whilst fully sampling the distribution would be preferable, since we have a metric, accurate sampling of every motif is not needed to characterize distributions over the graph at a coarse grained level.

\begin{figure}\centering
\includegraphics{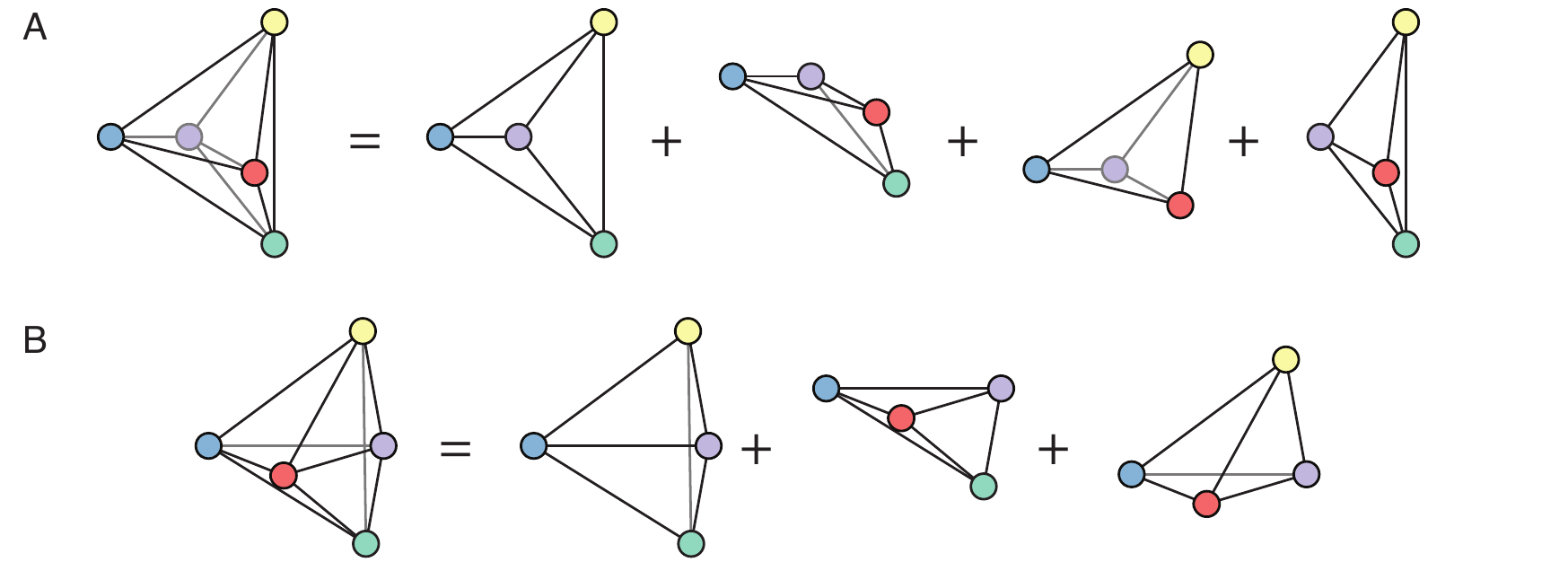}
\caption{\label{fig:Simplicial} Example of two different simplicial complexes with the same graph structure. 
(A) Simplicial complex created out of 4 simplices. (B) A different simplicial complex created from 3 simplices,
but with the same graph structure as (A).}
\end{figure}

In 2D it was equivalent to talk about the graph structure or the simplicial complex. The motifs were near-triangulations, meaning they were planar graphs with one non-triangular face. Taking the non-triangular face as the infinite face is enough to find an embedding, and so determine the simplices. In 3D, the graph structure alone is not enough to determine the simplicial structure, see Fig.~\ref{fig:Simplicial}. This means we could either represent the motifs as the full simplicial complex, or take a more coarse grained approach and just take the graph structure. Here we take the full simplicial complex.

\nolinenumbers
%\linenumbers

\subsection{Alpha complex}
One drawback of the Voronoi diagram is that every point in space (except at boundaries between regions) is assigned to a point. This means that points which are far away from each other in space can end up being neighbours in the Delaunay. For many of the systems considered here, connections between very far away points are unphysical, and such unphysical connections occur at the boundaries of the system, for instance the surface of a biofilm. Since we wish to study the bulk properties of materials, we do not include motifs which reside at boundaries. This is especially relevant if the boundary is artificial. For instance the star survey data only contains the closest stars to earth, which approximately fill out a sphere. The resulting boundary at the edge of this sphere has no physical meaning. 

To identify the motifs at the boundary, we work with the alpha-complex of the Delaunay. All tetrahedrons, $k$, in the Delaunay have a circumsphere with some radius, $r_k$. The alpha-complex is the simplicial complex made up of all tetrahedrons with circumsphere radius less than some parameter, $r_k < \alpha$~\cite{alpha_shape}.  This introduces a parameter $\alpha$ to be set, which we typically take to be $\alpha = 2 \times$~median~$r_i$. This should be chosen to eliminate only the unphysical tetrahedrons of the Delaunay tessellation, whilst retaining all else. For a particular system, say different biofilm experiments, we fix $\alpha$ across experiments for consistent analysis. Any motif which contains a tetrahedron with circumradius $r_k \geq \alpha$, i.e. too large to be in the alpha complex, is not counted as part of the topological distribution. We use this approach throughout this work, when data is not periodic and so has boundaries, following our previous approach in 2D~\cite{Skinner2021}.

Certain 3D systems of interest have few bulk points and mostly contain points on a surface. For these systems, simply ignoring all points on the boundary is not an appropriate option, yet the full Delaunay tessellation contains unphysical connections. In such a case we can still use the alpha complex,  which captures the topological neighborhood structure in a physically motivated way, but include the motifs on the boundary with large tetrahedrons removed in the motif distribution. The algorithms detailed here can run using these surface motifs, and we specify throughout when special cases of these algorithms arise due to such motifs. Whilst we have implemented this functionality, for consistency, we have not taken this approach for the systems here. Since the topological properties of the surface will typically differ from the bulk, for instance in the number of neighbors, by including surface information we are thus including global information about the ratio of surface to bulk points in the distribution which can be system size dependent. Instead, by only comparing bulk distributions throughout, we do not compare by system size but only by bulk topological properties.

\nolinenumbers
\begin{figure}\centering
\includegraphics{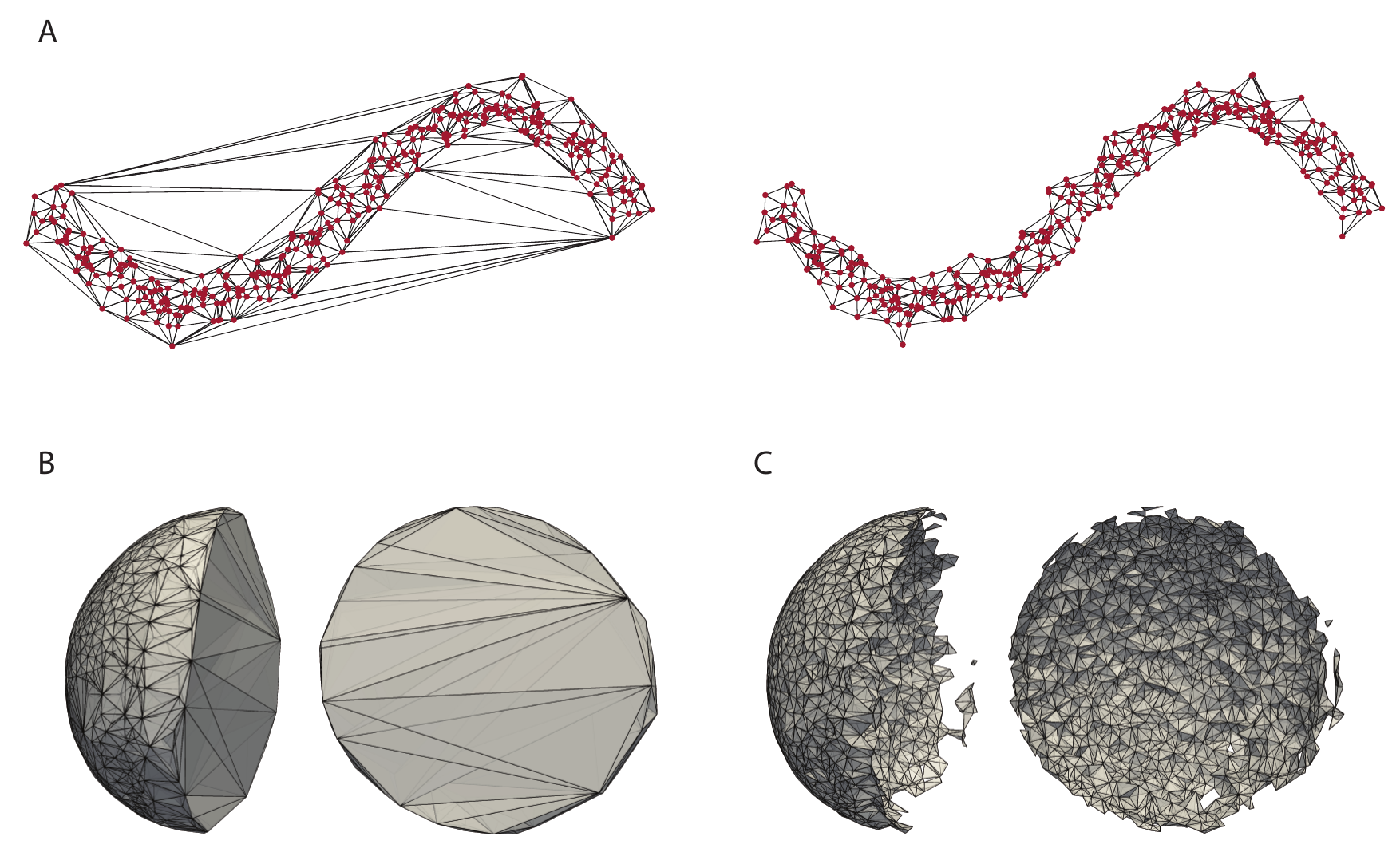}
\caption{\label{fig:Alpha}The alpha complex provides a way to extract a physically motivated simplicial complex from the Delaunay tessellation. (A) The Delaunay tessellation (black lines, left) of some points in $\mathbb{R}^2$ (red dots), contains unphysical connections between far apart points. Taking the alpha complex (black lines, right) by removing triangles with too large of a circumsphere, gives a physically motivated topological structure to work with. (B) For the developing zebrafish embryo at around 300 minutes post fertilization, we see that taking the Delaunay tessellation connects points that are far apart and physically unrelated (two views of the Delaunay tessellation are shown). (C) Taking the alpha complex gives a more physical topological object. Note that, we referred to this object as the ``Delaunay tessellation'' in the main text. Instead of showing the full Delaunay tessellation we only show the relevant part of it from which we extract motifs, or to be more precise the alpha complex.}
\end{figure}
%\linenumbers

\subsection{Comparing structures}
Given that we now characterize materials by the distribution of motifs, we need a way to compare these distributions. Methods such as Kullback-Leibler divergence or the Jensen-Shannon distance are able to compare distributions, however they are not aware of any structure on the space of motifs such as whether two motifs are similar in structure or not. This makes it difficult to compare distributions that have been undersampled~\cite{Skinner2021}, which is typically the case here. There is a natural metric on the space of motifs, which comes from the the idea of topological transitions. The Delaunay tessellation can only change through discrete topological transitions, and so the number of minimum number of flips to change one motif into another gives a notion of distance between motifs. These topological transitions, or flips, also gives a inherent graph structure, the flip graph, where two motifs are connected by an edge if they are one flip away from each other. The distance between motifs is then the minimum path length in this graph. Later, we introduce a spectral graph based distance, which can compare distributions in a way that is aware of the flip graph structure.

\section{Algorithmic implementation}
\subsection{Storing motifs}
Computationally, there is a need to store motifs in a concise representation allowing a set of motifs to be sorted into unique types, where two simplicial complexes are considered the same if some relabeling of the vertices makes them equal. In 2D this representation was the Weinberg vector~\cite{Skinner2021,LazarPRL}, but this required a planar graph, and hence cannot work in 3D where we use the simplicial complex rather than the graph representation (which would not be planar in any case). This task corresponds to finding a canonical labeling for the vertices of our simplicial complex.

\begin{figure}\centering
\includegraphics{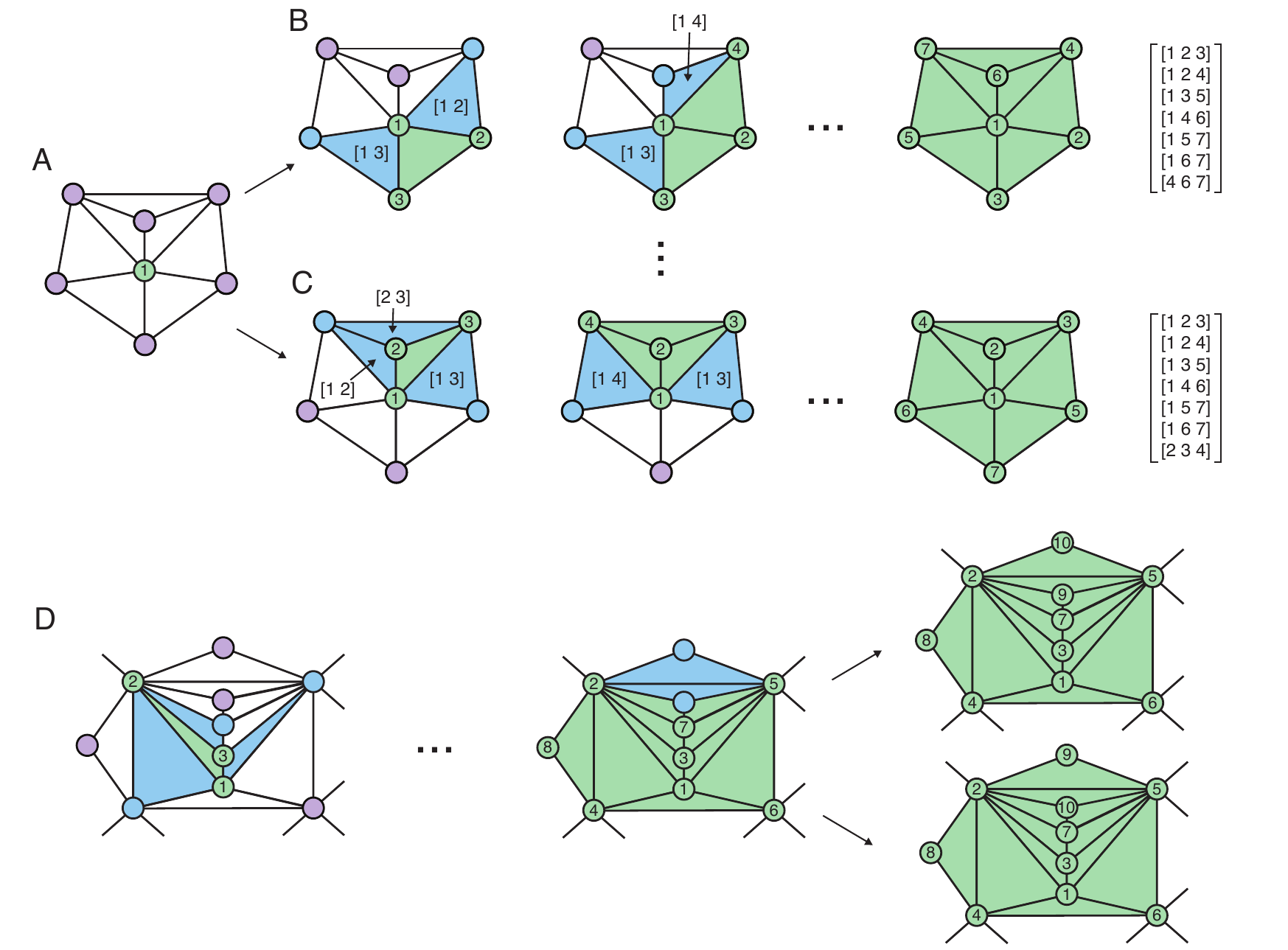}
\caption{\label{fig:CanonicalLabel} A worked example of canonically labeling a motif, and an example of the degenerate case. Vertices and simplices that are green have been labeled, those in blue could be labeled in the next step, and purple is to be labeled.
(A) A 2D motif with the central vertex labeled
as vertex 1. (B) A simplex adjoining 1 is chosen and labeled. There are now two vertices which could be labeled next with two corresponding simplices. Each of these simplices has two vertices labeled already, which are $(1 \; 3)$, and $(1\; 2)$. Since, lexicographically, $(1 \; 2) < (1 \; 3)$, we label the $4^{th}$ vertex as the one completing the simplex $(1 \; 2 \; 4)$. Next, there are two vertices and corresponding simplices which could be labeled. As $(1 \; 3) < (1 \; 4)$, the $5^{th}$ vertex is the one which completes the simplex $(1 \; 3 \; 5)$. This process continues until the motif is fully labeled, and the corresponding lexicographic ordered set of simplices is shown (right). (C) The procedure is repeated but with a different choice of initial labeling. The resulting labeling is different than (B), and results in a different set of simplices. The choice of labeling in (C) results in the minimum lexicographically ordered set of simplices, so this set of simplices serves as a topological identifier of this motif. (D) Part of a motif with an initial simplex labelled, which will result in a degenerate case. After several steps of the algorithm
the next simplex to be completed should be $(2 \; 5 \; 9)$, but there 
are two choices for vertex 9. This degeneracy is resolved by taking both
options and choosing at the end the minimum lexicographically
ordered set of simplices.}
\end{figure}

We start with the simplicial complex around a central vertex of some radius $r$. This consists of a number of $d$-simplices joined at their faces, where $d$ is the dimension, and a vertex specified as the central vertex. We always label the central vertex 1. Next we choose an arbitrary $d$-simplex that contains 1, and label the other vertices $2,\dots,d+1$, in some manner. We now construct a canonical labeling of the remaining vertices. Suppose we have labeled the first $k$ vertices, and we need to choose which vertex to label $k+1$. The candidates for the $k+1^{th}$ vertex are all vertices that lie on a $d$-simplex that has the other $d$ vertices labeled. Suppose that the other labeled vertices are $\{i_1,\dots,i_d\}, \{j_1,\dots,j_d\},\dots , \{l_1,\dots,l_d\}$ for different candidate vertices. We choose the vertex to label based on the lexicographic ordering, i.e. if \begin{equation} \{ i_1,\dots,i_d \} < \{j_1,\dots,j_d\} < \cdots < \{l_1,\dots,l_d \},\end{equation} then the $k+1^{th}$ vertex would be the one opposite to $\{i_1,\dots,i_d\}$. Once all vertices are labeled, take the lexicographically ordered set of simplices as the identifier of the graph. The initial simplex was picked arbitrarily, as was the labeling of the vertices $2, \dots, d+1$. To calculate the final canonical labeling, calculate the lexicographically ordered set of simplices for every possible initial labeling. Out of all of these simplices take the lexicographically first one, this is the canonical labeling of the graph, and two simplicial complexes will be isomorphic if and only if their canonically labeled set of simplices match. A worked example is shown in Fig. \ref{fig:CanonicalLabel}. In certain degenerate cases, these instructions are insufficient to fully label the simplicial complex and multiple possible labeling result from the same initial simpex labeling, see Fig.~\ref{fig:CanonicalLabel} for example and solution to this problem. In practice, for $r=1$ in 3D, such a degenerate case has never been observed.

\nolinenumbers
%\linenumbers

\subsection{Computing the flip graph}
\begin{figure}\centering
\includegraphics{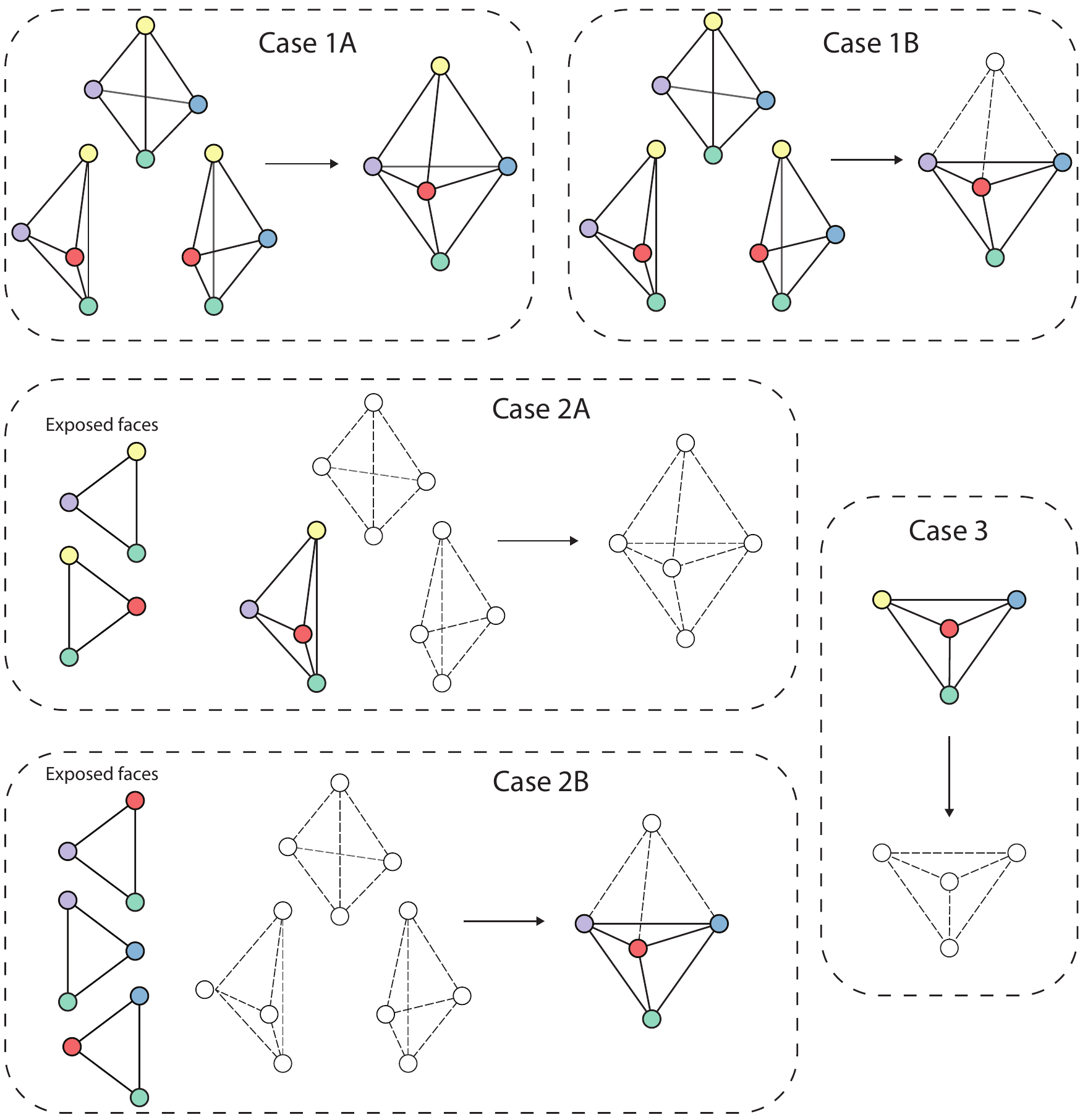}
\caption{ Possible cases showing how a flip alters a motif, shown in the direction of reducing the number of simplices. \label{fig:Cases}\textbf{Case 1A}: All post-flip vertices are also in the 
simplicial complex. This is just the full flip, which we note does not preserve the number of tetrahedrons. \textbf{Case 1B}: Post-flip, a vertex (yellow) is no longer in the simplicial complex. \textbf{Case 2A}: 4 vertices and 1 simplex are in the pre-flip simplicial complex, but no	simplices are in the post-flip simplicial complex. This can only happen to a local simplicial complex if certain faces are exposed as shown. \textbf{Case 2B}: 4 vertices and 0 simplices are in the pre-flip simplicial complex,	and 1 simplex is in the post-flip simplicial complex. This can only happen to a local simplicial complex if certain faces are exposed as shown. \textbf{Case 3}: a simplex starts in the pre-flip simplicial complex, but is not in the post-flip simplicial complex. This case only occurs when we are including surface motifs.}
\end{figure}

The flip graph is the graph where every vertex is a motif, and two motifs are connected by an edge if they are one flip, or topological transition, away from each other in the Delaunay triangulation. Note that, unlike in 2D, a flip in 3D does not preserve the number of tetrahedrons, Fig.~\ref{fig:Cases}. The alpha complex will change through flips similarly, but can also change when the circumsphere radius of a tetrahedron becomes too large, and that tetrahedron is no longer in the alpha complex. When including surface motifs, we will consider this tetrahedron removal step as a type of flip.

In order to compare distributions over the flip graph, we must calculate the flip graph numerically. To do so, we make use of the fact that a flip either increases or decreases the number of tetrahedrons by one. For each observed motif, we calculate all motifs that can be reached by a flip that decreases the number of tetrahedrons, which will account for every possible flip. We note that by flipping in this direction, the distance of any vertex to the central vertex can only decrease, and hence the vertices of the post-flip local simplicial complex will be a subset of the pre-flip local simplicial complex. Whilst the only allowed operation is a flip, how this affects the local simplicial complex depends on which vertices affected by the flip are elements of the local simplicial complex. This results in a number of different cases.

\textbf{Case 1}:
All vertices and simplices are in the initial simplicial complex. There are then two subcases, case 1A, Fig.~\ref{fig:Cases}, where all the post-flip vertices are in the simplicial complex, and case 1B, Fig.~\ref{fig:Cases}, where only 4 vertices are still in the post flip simplicial complex.

\nolinenumbers
%\linenumbers

\textbf{Case 2}:
Not all vertices are in the pre-flip simplicial complex. Case 2A, Fig.~\ref{fig:Cases}, has 4 vertices and 1 simplex in the pre-flip simplicial complex. Case 2B, Fig.~\ref{fig:Cases}, has 4 vertices and 0 simplices in the pre-flip simplicial complex. If a pre-flip simplicial complex had fewer than 4 vertices in it, it would not be affected by the flip.

\textbf{Case 3}:
This is the special case that we only consider when including surface motifs. In this case a tetrahedron in the simplicial complex is removed. This occurs when the circumradius has become too large, but note that when we compute the flip graph, we never actually calculate the circumradius. We simply know that such an operation is possible.

For every unique motif we calculate all motifs that are accessible by one flip that decreases the number of total simplices (Cases 1 and 2). In addition we calculate all motifs that can be reached by the reverse of flip 2A, in order to have a more connected flip graph, noting that this reverse flip can't increase the number of vertices in the local motif. If we are including surface motifs, then we also calculate motifs that are accessible through Case 3. From this newly enlarged set of motifs, we calculate which edges exist by taking all possible flips for every motif. Typically this results in a flip graph with over 95\% of the original motifs in the largest connected component. If this flip graph proves too large for practical computations, we reduce it in size by calculating the page rank of each vertex, a measure of graph centrality. If the page rank of a motif is below some threshold, and that motif is not one of the original observed motifs, we remove it. By adjusting this threshold, one can typically reduce the graph size by a factor of 5 or so, with over 90\% of original motifs still in the largest connected component of the reduced graph. 

\section{Choice of distance}
\subsection{Approximating the earth mover's distance}
Recall that given material $A$ and material $B$, computing local motifs yields probability distributions $\rho_A$, and $\rho_B$ over the discrete set of possible motifs. We have the additional structure of the flip graph, giving us a metric on the space of motifs, and we would like our comparison of $\rho_A$ and $\rho_B$ to be aware of this structure. One such comparison that is graph aware, is to use the Wasserstein or earth mover's distance to give a notion of distance between materials~\cite{Skinner2021}. Computing the distance numerically is equivalent to solving a minimum cost flow problem. 

Specifically, consider the flip graph to be directed with edges oriented arbitrarily. Then,  let $J_e$ be the flow along the $e^{th}$ directed edge. We also define the incidence matrix
\begin{equation}
 D_{ev} = \left\{ \begin{array}{c} 1 \text{ if } \exists \; w, \; e = (v,w), \\
-1 \text{ if } \exists \; w, \; e = (w,v), \\
0 \text{ else}. \end{array} \right. \end{equation}
The distance phrased as a minimum cost flow problem is then
\begin{equation} d_\text{TEM}(A,B) = \min || J ||_1 \quad \text{subject to  } D^T J = \rho_A - \rho_B. \end{equation}
While there is nothing preventing us from using this distance, it becomes extremely expensive to compute for many of the datasets used here. In particular, the size of the flip graph in 3D perhaps a factor of 5-10 times larger than in 2D, making the minimum cost flow problem far more expensive. We have some discretion in how large the flip graph is, for instance by only including motifs that have been observed more than $n$ times for some $n$, or by setting the page-rank threshold to be higher, and hence restricting the graph size. However, we can get a huge reduction in computational time by changing the distance we use, whilst not compromising on our goal of having a distance that is aware of the flip graph structure. We introduce this distance now.

Following Ref.~\cite{DiscreteEMD}, we start by rewriting our expression for the earth mover's distance by using a Helmholtz-like decompositon of the flow as
\begin{equation} 
J = J_0 + Df, 
\end{equation}
where $D^T J_0 = 0$ and $f$ is arbitrary, the decomposition follows from rank nullity. The constraint then becomes
\begin{equation} D^T D f = \rho_A - \rho_B,\end{equation}
in which we recognise $L = D^TD$ as the discrete graph laplacian. Taking the pseudo-inverse, $f = L^+ (\rho_A - \rho_B)$, the minimization problem is
\begin{equation} d_\text{TEM}(A,B) = \min || J_0 + D L^+ (\rho_A - \rho_B) ||_1 \text{ subject to } D^T J_0 = 0.\end{equation}
Since $J_0$ is some element in the kernel $D^T x = 0$, we can write $J_0 = \sum_{i=1}^N a_i K_i$, where the $K_i$'s form a basis for the kernel and the $a_i$'s are coefficients. Formally, the minimization is
\begin{equation} d_\text{TEM}(A,B) = \min_{a_i} || \sum_{i=1}^N a_i K_i + D L^+ (\rho_A - \rho_B) ||_1, \end{equation}
which now has no constraints. It is also possible to define a family of distances using a restricted sum on the kernel vectors~\cite{DiscreteEMD},
\begin{equation} d_{M}(A,B) = \min_{a_i} || \sum_{i=1}^M a_i K_i + D L^+ (\rho_A - \rho_B) ||_1, \end{equation}
for $0\leq M < N$. For graphs that come from the discretization of a smooth manifold, one can observe spectral convergence and thus approximate $d_{TW} \approx d_{M}$ for $M \ll N$~\cite{DiscreteEMD}. The flip graph does not arise from a smooth manifold, and so we observe only linear convergence.  A large number of kernel vectors would therefore be needed for a good approximation to the earth mover's distance, saving minimal computational time. We consider instead the distance defined by taking
$M=0$, hence,
\begin{equation} d_\text{Diff}(A,B) = || D L^+ (\rho_A - \rho_B) ||_1. \end{equation}
This defines a distance that can be calculated simply by solving a linear system of equations. While this may not be a good approximation to the earth mover's distance, we show that, like the earth mover's distance, it has a natural physical interpretation.

The diffusion equation for a scalar field $\phi$ over a graph with sources and sinks of strength $\rho_A - \rho_B$, is
\begin{equation} \frac{\mathrm{d} \phi}{\mathrm{d} t} + L \phi = \rho_A - \rho_B, \end{equation}
with diffusion constant set to 1. The steady state of this equation, has
\begin{equation} \phi = L^+ (\rho_A - \rho_B) + const.,\end{equation}
and the net flow rate is given by
\begin{equation} J = D \phi. \end{equation}
Therefore the $L_1$ cost of the flow is 
\begin{equation} d_\text{Diff}(A,B) = ||D\phi||_1 = || D L^+ (\rho_A - \rho_B) ||_1. \end{equation}
For optimal transport, the natural interpretation was that it was the minimum number of flips needed to make distribution $A$ look like distribution $B$. This could also be interpreted physically, as the energetic distance between distributions when each flip is associated with crossing an energy barrier as with epithelial cells. The distance $d_\text{Diff}$ also has a natural interpretation; it is the number of flips needed to make distribution $A$ look like distribution $B$, but without any control over which flip occurs. The probability mass undergoes diffusion on the graph with sources and sinks, and the rate of flipping corresponds to the $L_1$ norm of the flow. 

In many ways the diffusive distance $d_\text{Diff}(A,B)$ is more physically appealing than the earth mover's distance, $d_\text{TEM}(A,B)$, random flips are more physically realizable than a targeted minimization over all possible flows. The computation is orders of magnitude faster and involves solving a sparse linear system, for which optimized iterative solvers exist. In any case, as long as the distance used is aware of the underlying metric on motifs, the key conceptual ideas of the method are utilized.

\begin{figure}\centering
\includegraphics{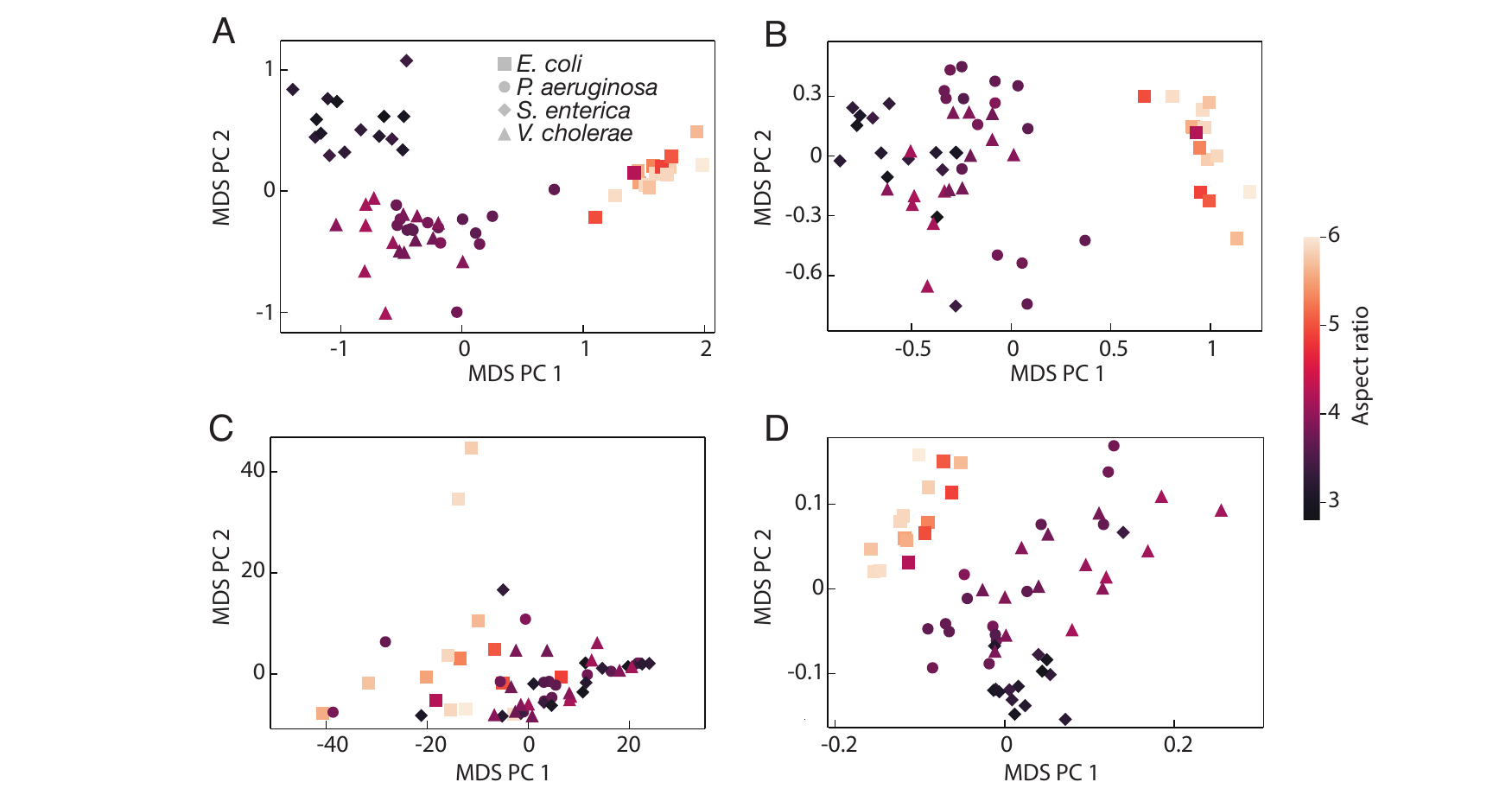}
\caption{ \label{fig:distComp}Comparison of the diffusion based distance $d_\text{Diff}$, the earth mover's distance $d_\text{TEM}$, and the bottleneck distance, $d_\text{BN}$, as applied to bacterial biofilm data. (A) 2D embedding of the diffusion distance colored by average cellular aspect ratio, as appears in Fig.~1D. (B) Embedding of the earth mover's distance is broadly similar to the diffusion distance embedding, but does not separate \emph{S. enterica} as clearly from \emph{P. aeruginosa} and \emph{V. cholerae} biofilms.  (C) The Bottleneck distance embedding shows no discernible structure and fails to separate out different species of biofilm. (D) Embedding based on the Jensen-Shannon (JS) distance is slightly worse than the earth mover's distance and is unable to separate fully \emph{S. enterica} from the other species, but outperforms the bottleneck distance.}
\end{figure}

Whilst the earth mover's distance may be too expensive to use for certain computations done here, we were able to compute the pairwise earth mover's distance between all pairs of biofilm experiments, enabling us to compare directly to the diffusion distance. We find that, whilst both distances result in a similar embedding, there is actually more structure in the diffusion distance embedding, Fig.~\ref{fig:distComp}. From now on, we therefore use the diffusion distance.

\nolinenumbers
%\linenumbers

\subsection{Comparison to Bottleneck distance}
Whilst the framework introduced here takes a topological approach to analyze data, our approach is distinct from the field of topological data analysis (TDA). As originally developed, TDA seeks to understand the ``shape'' of a high dimensional data manifold by characterizing it topologically~\cite{carlsson2009topology}. Typically, the starting point is the Vietoris-Rips complex, a topological object somewhat related to the Voronoi diagram, but containing more detailed information, and from this further features are computed, such as the number of $n$-dimensional ``holes''~\cite{carlsson2009topology}. These resulting features can be summarized using the persistence diagram which, ultimately, represents key topological features as a set of points in $\mathbb{R}^2$~\cite{carlsson2009topology}. Two objects, say different biofilms, can then be compared by comparing their persistence diagrams.  a A common way to compare persistence diagrams is to use the bottleneck distance, which, as each persistence diagram is a collection of points in $\mathbb{R}^2$, is essentially the earth mover's distance between these two collections of points, but accounting for an unequal number of points~\cite{edelsbrunner2008persistent}.

We use the computational framework of Ref.~\cite{Cufar2020}, to compute the pairwise bottleneck distance between different biofilms using their one dimensional persistence diagram. From this distance matrix, we embedded the points into Euclidean space with MDS. This embedding does not reveal any of the structure of the data, Fig.~\ref{fig:distComp}C, and whilst the \emph{E. coli} form a somewhat distinct cluster, different species are all intermixed. Moreover, this computation was orders of magnitude slower than computing the diffusion distance. This is not surprising as the bottleneck distance is not physically motivated for the 3D systems studied here. Specifically, whilst understanding the `shape'' of a data manifold is a key challenge for high dimensional data, for the 3D data we work with, the local structure is more important physically than larger scale features such as holes.

\subsection{Comparison to Jensen-Shannon distance}
A widely used class of distances compare distributions based on quantities related to their relative entropies, such as the the Kullback-Leibler divergence or the Jensen-Shannon (JS) distance. Such distances arise naturally in information theory, but are unaware of any metric, in particular they do not use the graph structure on the space of motifs. The JS distance between discrete distributions $p$ and $q$ is
\begin{equation}
    d_{JS}(p,q)^2 = \frac{1}{2} \sum_i \  p_i \log \frac{p_i}{z_i} + q_i \log \frac{q_i}{z_i}, \qquad z_i = \frac{1}{2}(p_i + q_i),
\end{equation}
where we note that this distance only uses frequencies of states and not any structural information about the underlying space.
In 2D, we previously found that JS distances performed worse than optimal transport, particularly for subsampled distributions~\cite{Skinner2021}. Here, we find that the TDD outperforms JS, Fig.~\ref{fig:distComp}D, although JS still performs significantly better than the TDA bottleneck distance whilst requiring orders of magnitude less computational cost. We also repeated the biofilm bootstrapping calculation, that will be described in Section~\ref{sec:Bootstrap}, using the JS distance. We found that whilst the TDD can distinguish \emph{V. cholarea} and \emph{P. aeruginosa} at $p<0.01$, the JS distance can not, although it can distinguish all other pairs at $p<0.01$. The bootstrap calculation was too expensive to perform with the earth mover's or bottleneck distances.

\section{Menger curvature}
Our topological diffusion distance creates a metric space where points are distributions over the space of motifs. Remarkably, the field of distance geometry allows us to build geometric concepts from such an abstract metric space alone~\cite{Liberti2016}. For instance, a well known and arguably the earliest result in distance geometry is Hero's formula for the area of a triangle; allowing the computation of the area from only the distances between points. Later work by Arthur Cayley and Karl Menger established conditions for an abstract (semi-)metric space to be equivalent to $\mathbb{R}^n$ for some $n$~\cite{Liberti2016}. Here, we introduce a geometric notion of curvature along a 1D path in topological space, which could be extended to define curvature for a surface or general manifold.

Consider a topological distribution that changes with some parameter, for instance time, $\rho_t$. We would expect the MDS embedding of such a process to recover the temporal ordering, as indeed we find in Fig. 2D (main text) for zebrafish development. However, the embedding may take a highly curved path, as was seen in Fig. 2D, or follow a straight line, as was found in two dimensions for the development of a fly wing~\cite{Skinner2021}. In Euclidean space, a straight line is the optimal way to move between two points, whereas a curved path is longer. This gives us an intuition in our topological space that a system which takes a straight path acts to minimize the number of topological flips, whereas a system with a curved path performs more flips, and hence more rearranging, than is necessary. In making this intuition precise, we wish to avoid the distorting effect that a low dimensional embedding may have on a trajectory, and so we want to have a notion of how curved the path is at a point, independent of the embedding.

To start, we consider the notion of a straight line, or geodesic, under the $d_{Diff}$ distance. A path $\rho_t$ between $\rho_0$ and $\rho_1$ is a geodesic if 
\begin{equation}
    \sum_{i=0}^n d_{Diff}(\rho_{t_i},\rho_{t_{i-1}}) =  d_{Diff}(\rho_0,\rho_1),
\end{equation}
for all $0=t_0 < \cdots < t_n = 1$.
Not all paths are geodesics, but as with  the space of Wasserstein 1 ($W_1$) or earth mover's distances~\cite{solomon2016continuous}, geodesics are not unique in the space of $d_{Diff}$. 

For instance, an example of an optimal path would be $\rho_t = \rho_0 + t(\rho_1 - \rho_0)$, where given any finite $0=t_0 < t_1 < \cdots < t_n = 1$,
\begin{equation}
    \sum_{i=1}^n d_{Diff}(\rho_{t_i},\rho_{t_{i-1}}) = \sum_{i=1}^n || D L^{\dagger} (\rho_{t_i} - \rho_{t_{i-1}}) ||_1 =
    \sum_{i=1}^n || D L^{\dagger} (t_i - t_{i-1})(\rho_1-\rho_0) ||_1 = d_{Diff}(\rho_0,\rho_1).
\end{equation}
However, this path corresponds to phase separated growth, at time $t$, a fraction $t$ of the system is in the $\rho_1$ phase, and a fraction $1-t$ is in the $\rho_2$ phase. To identify the most physical continuous path between $\rho_0$ and $\rho_1$ we will need additional structure beyond the $d_{Diff}$ distance.

Following our previous work in 2D, and building on the work of Ref.~\cite{solomon2016continuous}, we can find the most natural geodesic by choosing the path between $\rho_0$ and $\rho_1$ that additionally minimizes a dissipation like term.  
\begin{align}
    &  \inf_{J(t,e), p(v,t) \geq 0} \int_0^1 \sum_{e = (u,v)} \frac{J(t,e)^2}{2} \left( \frac{1}{p(t,u)} + \frac{1}{p(t,v)} \right) \d t \\\notag
    &\text{such that}, \quad \sum_u p(t,u) = 1, \ p(0,v) = \rho_0(v),\ p(1,v) = \rho_1(v), \\\notag
    &\frac{\d}{\d t} p  = D^\top J, \qquad \sum_{i=1}^{n+1} d_{Diff}(p(t_i,\cdot),p(t_{i+1},\cdot)) = d_{Diff} (\rho_0,\rho_1),
\end{align}
where we are minimizing the square of a current $J(t,e)$ along an edge $e$ divided by probability mass, which has the interpretation of a squared velocity multiplied by a probability mass; a dissipation like term~\cite{solomon2016continuous}. The resulting infimum $p(t,\cdot)$ gives us the unique dissipation minimizing geodesic path. The final condition ensures that any path that is taken is still a geodesic under the $d_{Diff}$ metric, and we will prove later that condition is automatically enforced by the local minimization and does not need to be additionally imposed.

The interpretation of the (unique) infimum, $J$ and $p$, is that they describe a unique geodesic between $\rho_0$ and $\rho_1$. One can interpret this geodesic as the minimum dissipation path to move mass from $\rho_0$ to $\rho_1$, and it results in mass being locally transported across the flip graph, rather than being moved discontinuously~\cite{solomon2016continuous}.

%%%%%%%%%%%%%%%%%%%%%
Instead of choosing a special path out of many geodesics of $d_{Diff}$, these geodesics arise naturally from an alternate distance $\bar{W}$ between two distributions~\cite{solomon2016continuous}, namely
\begin{align}\label{eq:Wbar}
    [\bar{W}(\rho_0,\rho_1)]^2 =&  \inf_{J(t,e), p(v,t) \geq 0} \int_0^1 \sum_{e = (u,v)} \frac{J(t,e)^2}{2} \left( \frac{1}{p(t,u)} + \frac{1}{p(t,v)} \right) \d t \\\notag
    &\text{such that}, \quad \sum_u p(t,u) = 1, \ p(0,v) = \rho_0(v),\ p(1,v) = \rho_1(v), \\\notag
    &\qquad   \frac{\d}{\d t} p  = D^\top J.
\end{align}
Computing the $\bar{W}$ distance requires solving a second-order conic system, which is even more expensive than the $W_1$ computation, and making it impractical for the size of systems we work with. However, we will show that we can use the properties of this distance to define a notion of local curvature, relating how close an observed path comes to the dissipation minimizing geodesic, without ever needing to explicitly calculate $\bar{W}$.

\begin{figure}\centering
\includegraphics{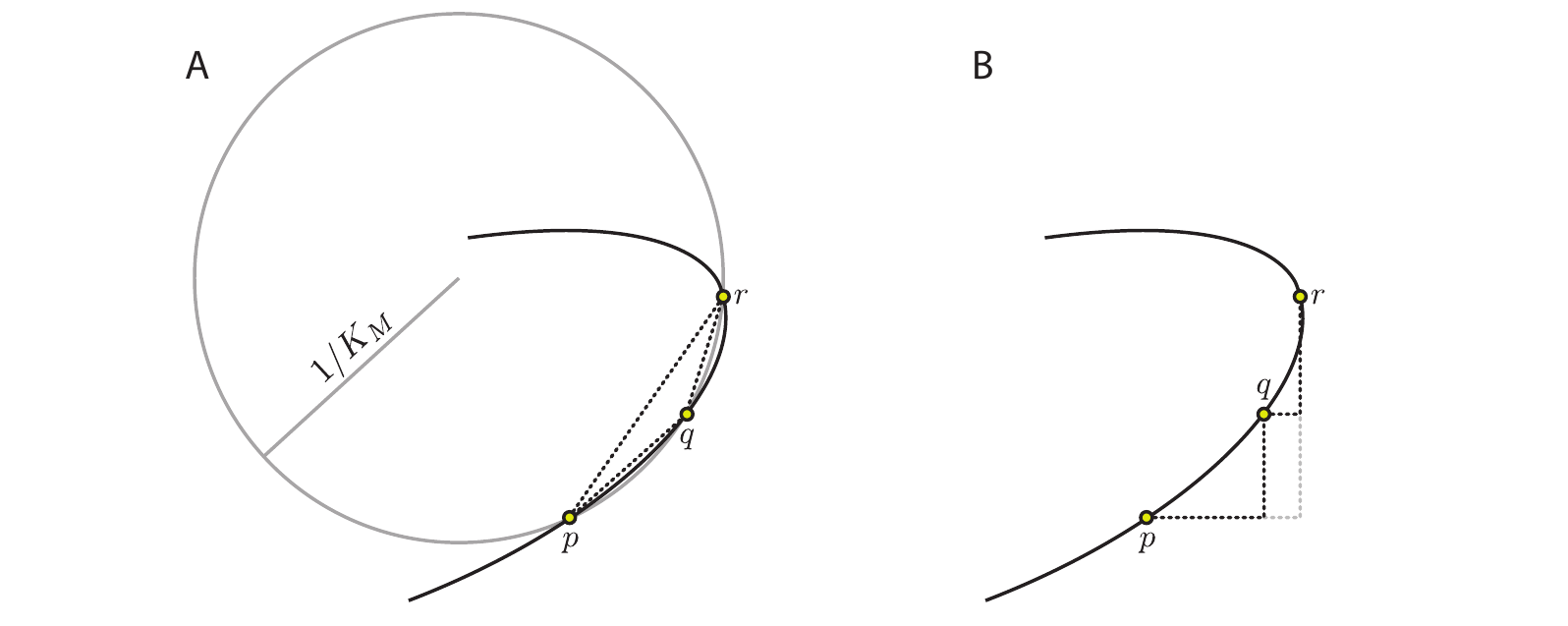}
\caption{\label{fig:Menger}Menger curvature for a plane curve with three points $p$, $q$, $r$. (A) The points $p$, $q$, $r$ on some curve form a triangle in the plane, with corresponding circumcircle shown in gray. The radius of this circumcircle can be found by only knowing the pairwise distances between the points, using a well known formula, and Menger curvature $K_M(p,q,r)$ is defined to be the reciprocal of this radius. Since this formula, Eq.~\eqref{eq:Menger}, only uses pairwise distances, it can be extended to give a concept of curvature in any metric space.  (B) With the $L_1$ metric in the plane, the points $p$, $q$, $r$ lie on a straight line, as the path going through all three (black dashed line) is just as long as the direct path between $p$ and $r$ (gray dashed line). In this case, the Menger curvature would be zero. }
\end{figure}

Generally speaking, consider that we observe a path $\rho_t$, in some space where a unique geodesic arises from some metric $d$. We wish to to quantify how close our path is to being a geodesic. To do so, we draw on motivation from Euclidean geometry, even though the following definition applies for any metric space. We define the Menger curvature~\cite{saucan_samal_jost_2021} of points $p,q,r$ as follows, 
\begin{equation}\label{eq:Menger}
    K_M(p,q,r) = \frac{\sqrt{[pq + qr + rp][-pq + qr + rp][pq - qr + rp][pq + qr - rp]}}{pq\cdot qr \cdot rp}
\end{equation}
for a general metric $d:X\times X \to \mathbb{R}$, and $pq := d(p,q)$, etc.. This is also the inverse of the radius of curvature of a triangle with side lengths $d(p,q),$  $d(q,r)$, and $d(r,p)$, see Fig.~\ref{fig:Menger}. The curvature quantifies the extent to which, locally, the path does not take the shortest path under the metric $d$.

\nolinenumbers
%\linenumbers

To locally define a curvature, for our curve $\rho_t$, we would like to take $K_M(\rho_t,\rho_{t+\Delta t},\rho_{t-\Delta t})$ in the limit $\Delta t \to 0$ and this the curvature of the path at $\rho_t$. For case of the $L_1$ metric in the plane, this will result in almost all points having zero curvature, Fig.~\ref{fig:Menger}B (a measure zero set may have a diverging curvature). Similarly, with $d_{Diff}$ based on an underlying $L_1$ metric, the curvature $K_M$ will be almost everywhere zero, as locally many paths will be geodesics. However, using the $\bar{W}$ distance, which gives rise to a unique geodesic, will generally result in a  non-vanishing curvature.  We therefore seek to use $\bar{W}$ to compute the curvature along a curve, and specifically we are using this curvature as a measure of how much a given path differs from the dissipation minimizing path under $d_{Diff}$ which locally is governed by the metric $\bar{W}$. As we only need to compute $\bar{W}$ between asymptotically similar distributions $\rho_{t}$,$\rho_{t + \Delta t}$,$\rho_{t-\Delta t}$, we need not solve a second order conic system, and the computational cost is only slightly more expensive than computing the $d_{Diff}$ distance.

\subsection{Curvature perturbation problem}
Formally, we seek an asymptotic solution of the $\bar{W}$ distance between distributions $p = p_0 + \epsilon p_1 + \epsilon^2 p_2 + \dots$, and $r = p_0 + \epsilon r_1 + \epsilon^2 r_2 + \dots$, where $\epsilon$ is small, and by construction $p$ and $r$ converge to the same distribution in the $\epsilon \to 0$ limit. We have an expansion of the form $J = \epsilon J_1 + \epsilon^2 J_2 + \dots$, and $q = p_0 + \epsilon q_1 + \epsilon^2 q_2 + \dots$, and plugging these into Eq.~\eqref{eq:Wbar} gets the following hierarchy,
\begin{align}\notag
\bar{W}^2 =& \epsilon^2 I_0(J_1,q_1,p_0,p_1,r_1) + \epsilon^3 I_1(J_1,J_2,q_1,q_2,p_0,p_1,p_2,r_1,r_2) \\ &\ + \epsilon^4 I_2(J_1,J_2,J_3,q_1,q_2,q_3,p_0,p_1,p_2,p_3,r_1,r_2,r_3) + \dots,
\end{align}
where each $I_j$ is a minimization problem, and higher order $I_j$'s depend on variables that were fully or partially determined by lower order ones. We will need to solve up to $I_2$, and also note here that 
\begin{equation}
    \bar{W} = \epsilon \sqrt{I_0} + \frac{\epsilon^2}{2} \frac{I_1}{\sqrt{I_0}} + \epsilon^2 \left[ \frac{1}{2} \frac{I_2}{\sqrt{I_0}} - \frac{1}{8} \frac{I_1^2}{I_0^{3/2}} \right] + \dots,
\end{equation}
which we will make use of when computing the curvature later.

\subsubsection{Zeroth order}
At the lowest order, the problem we are trying to solve becomes
\begin{align}\label{eq:I0}
    I_0 =&  \inf_{J_1, q_1} \int_0^t \sum_{e = (u,v)} \frac{J_1(t,e)^2}{2} \left( \frac{1}{p_0(u)} + \frac{1}{p_0(v)} \right) \d t \\\notag
    &\text{such that}, \quad q_1|_{t=0} = p_1,\ q_1|_{t=1} = r_1, \\\notag
    &\qquad    D^\top J_1 = \dot{q}_1,
\end{align}
where no  requirement that $q_1$ must be non-negative exists.  Now that we are free from all positivity constraints, we can find the minimum by using Lagrange multipliers
\begin{align}
    \mathcal{L} =&  \int_0^t \sum_{e = (u,v)} \frac{J_1(t,e)^2}{2} \left( \frac{1}{p_0(u)} + \frac{1}{p_0(v)} \right) + \sum_{w} \lambda_0(t,w) \left(\sum_{e = (u,v)} D_{ew} J_1(t,e) -\dot{q}_1(t,w)\right) \d t,
\end{align}
where a minimizing solution satisfies
\begin{align}
    \frac{\delta \mathcal{L}}{\delta J_1} =&   J_1(t,e) \left( \frac{1}{p_0(u)} + \frac{1}{p_0(v)} \right) + \sum_{w} D_{ew} \lambda_0(t,w)     =0, \\ 
    \frac{\delta \mathcal{L}}{\delta q_1} =& \dot{\lambda_0}(t,w) = 0.
\end{align}
From this we can immediately identify that the Lagrange multiplier $\lambda_0$ does not depend on time, thus neither does $J_1$, and so $q_1$ is linear in time, and moreover we can deduce $q_1$ from the initial and final conditions. We therefore have that
\begin{align}
    q_1 &= (1-t) p_1 +  tr_1, \\
    J_1(e) \left( \frac{1}{p_0(u)} + \frac{1}{p_0(v)} \right) + \sum_{w} D_{ew} \lambda_0(w)     &= 0, \\ 
    \sum_{e = (u,v)} D_{ew} J_1(e) &= r_1(w) - p_1(w).
\end{align}
From now on, understanding $D = (D_{ev})$ to only have indices that run over edges $e=(u,v), u<v$, and defining 
\begin{equation}
    \Lambda_{e e'} = \delta_{e e'} \left( \frac{1}{p_0(u)} + \frac{1}{p_0(v)} \right),
\end{equation}
we have that
\begin{equation}
    \Lambda J_1 + D \lambda_0 = 0 \implies D^\top J_1 + D^\top \Lambda^{-1} D \lambda_0 = 0 \implies L \lambda_0 = -(r_1 -p_1),
\end{equation}
where $L = D^\top \Lambda D$, and whilst the solution of $L \lambda_0 = p_1 - r_1$ is not unique, it will always lead to the same $J_1$, which is unique. Overall,
\begin{equation}
    I_0 = - \lambda_0^\top (r_1 - p_1), \quad \text{where}\ L\lambda_0 = -(r_1-p_1)
\end{equation}

\subsubsection{First order}
At the next order, we are trying to solve
\begin{align}
    I_1 =&  \inf_{J^i_2, q^i_2} \int_0^1 \sum_{e = (u,v)}\  J_2(t,e)J_1(e) \left( \frac{1}{p_0(u)} + \frac{1}{p_0(v)} \right) + \frac{J_1(e)^2}{2} \left( -\frac{q_1(t,u)}{p_0(u)^2} - \frac{q_1(t,u)}{p_0(v)^2} \right) \d t \\\notag
    &\text{such that}, \quad q_2^0 = p_2,\ q_2^k = r_2, \quad    D^\top J_2 = \dot{q_2},
\end{align}
which in light of the known form of $J_1$, $q_1$ we can rewrite as
\begin{align}
    I_1 =&  \inf_{J_2, q_2}  \sum_{e = (u,v)} \left[\int_0^1 J_2(t,e)\d t\right] \Lambda_{ee}J_1(e) -\sum_{e = (u,v)} \frac{J_1(e)^2}{4} \left( \frac{r_1(u) + p_1(u)}{p_0(u)^2} + \frac{r_1(v) + p_1(v)}{p_0(v)^2} \right)   \\\notag
    &\text{such that},   D^\top \left[\int_0^1 J_2(t,e)\d t\right] = r_2 - p_2,
\end{align}
where we can rewrite part of the objective as
\begin{align}
     \sum_{e = (u,v)} \left[\int_0^1 J_2(t,e)\d t\right] \Lambda_{ee}J_1(e) = -\sum_{e = (u,v),w} \left[\int_0^1 J_2(t,e)\d t\right] D_{ew} \lambda_0(w) = - \sum_w (r_2(w) - p_2(w)) \lambda_0(w)
\end{align}
so that in total
\begin{equation}
    I_1 = - \sum_w (r_2(w) - p_2(w)) \lambda_0(w) -\sum_{e = (u,v)} \frac{J_1(e)^2}{4} \left( \frac{r_1(u) + p_1(u)}{p_0(u)^2} + \frac{r_1(v) + p_1(v)}{p_0(v)^2} \right)
\end{equation}
and at this level, we need not find $J_2$ nor $q_2$ explicitly. 

\subsubsection{Second order}
At second order we are solving
\begin{align}
    I_2 =&  \inf_{J_2, q_2, J_3, q_3} \int_0^1 \sum_{e = (u,v)} \left[J_3(t,e)J_1(e) + \frac{J_2(t,e)^2}{2}\right] \left( \frac{1}{p_0(u)} + \frac{1}{p_0(v)} \right) + J_1 (e) J_2(t,e) \left( -\frac{q_1(t,u)}{p_0(u)^2} - \frac{q_1(t,v)}{p_0(v)^2} \right)  \\\notag
    & \qquad + \frac{J_1(e)^2}{2} \left(\frac{q_1(t,u)^2/2 - p_0(u) q_2(t,u)}{p_0(u)^3} + \frac{q_1(t,v)^2/2 - p_0(v)q_2(t,v)}{p_0(v)^3} \right) \d t \\\notag
    &\text{such that}, \   q_2|_{t=0} = p_2,\ q_2|_{t=1} = r_2, \ D^\top J_2 = \dot{q}_2,\ q_3|_{t=0} = p_3,\ q_3|_{t=1} = r_3, \ D^\top J_3 = \dot{q}_3.
\end{align}
Firstly, by the same logic as at first order, we know the term involving $J_3$ as
\begin{align}
     \sum_{e = (u,v)} \left[\int_0^1 J_3(t,e)\d t\right] \Lambda_{ee}J_1(e) =  - \sum_w (r_3(w) - p_3(w)) \lambda_0(w)
\end{align}
and moreover, we will see later that this term cancels in the curvature computation. We can also compute
\begin{align}
    \chi_3 &= \int_0^1 \sum_{e = (u,v)} 
     \frac{J_1(e)^2}{4} \left(\frac{q_1(t,u)^2 }{p_0(u)^3} + \frac{q_1(t,v)^2}{p_0(v)^3} \right) \d t \\ \notag 
     &= \frac{J_1(e)^2}{12} \left(\frac{r_1(u)^2 + r_1(u)p_1(u) + p_1(u)^2 }{p_0(u)^3} + \frac{r_1(u)^2 + r_1(u)p_1(u) + p_1(u)^2}{p_0(v)^3} \right).
\end{align}
We can write the remaining minimization problem with constraints as Lagrange multipliers as 
\begin{align}
     \mathcal{L} &=  \int_0^1 \sum_{e = (u,v)}  \frac{J_2(t,e)^2}{2} \left( \frac{1}{p_0(u)} + \frac{1}{p_0(v)} \right) + J_1 (e) J_2(t,e) \left( -\frac{q_1(t,u)}{p_0(u)^2} - \frac{q_1(t,v)}{p_0(v)^2} \right)  \\\notag
    &\qquad + \frac{J_1(e)^2}{2} \left(\frac{ -  q_2(t,u)}{p_0(u)^2} + \frac{ - q_2(t,v)}{p_0(v)^2} \right) + \sum_w \lambda_1(w) \left( \sum_e D_{ew} J_2(t,e) - \dot{q}_2(w) \right)\d t,
\end{align}
or

\begin{align}
     \mathcal{L} &= \int_0^1 \frac{1}{2} J_2^\top \Lambda J_2 + J_1^\top \tilde{\Lambda} J_2 + \Gamma^\top  q_2  +  \lambda_1^\top\left( D^\top J_2 - \dot{q}_2 \right)\d t,
\end{align}
in concise matrix notation, with
\begin{equation}
    \Gamma(u) =  \sum_{v} \frac{J_1(e=(u,v))^2}{2p_0(u)^2} + \sum_{v} \frac{J_1(e=(v,u))^2}{2p_0(u)^2},
\end{equation}
where the sum is taken over edges that exist without double counting. We find the following,
\begin{align}
    \frac{\delta \mathcal{L}}{\delta J_2} =& \Lambda J_2 + \tilde{\Lambda}J_1 + D \lambda_1     =0, \\ 
    \frac{\delta \mathcal{L}}{\delta q_2} =& \Gamma + \dot{\lambda_1} = 0,
\end{align}
from which we deduce that $J_2$ is linear in time, and $q_2$ is quadratic. Calling
\begin{align}
    \tilde{\Lambda}J_1 &= J_1(e) \left( -\frac{q_1(t,u)}{p_0(u)^2} - \frac{q_1(t,v)}{p_0(v)^2} \right) \\\notag 
    &= J_1(e) \left( -\frac{(1-t)p_1(u) + t r_1(u)}{p_0(u)^2} - \frac{(1-t)p_1(v) + t r_1(v)}{p_0(v)^2} \right) \\
    &= -(\chi_1 + t \chi_2),
\end{align}
and $q_2 = (1-t)p_2 + t\, r_2 + t(1-t) s$, we have that
\begin{align}
    J_2 &=  \Lambda^{-1} (\chi_1 + t \chi_2) + t \, \Lambda^{-1} D \Gamma + \Lambda^{-1}Dc
\end{align}
so
\begin{equation}
    D^\top\Lambda^{-1} (\chi_1 + t \chi_2) + t \, L \Gamma + L c = r_2 - p_2 + (1-2t)s
\end{equation}
and since this holds for all $t$, this implies that 
\begin{align}
    D^\top\Lambda^{-1} (\chi_1 )  + L c =& r_2 - p_2 + s \\ \notag
    D^\top\Lambda^{-1} ( \chi_2) + L \Gamma  =& -2s,
\end{align}
thus we immediately know $s$ and can find $c$ through one linear solve. In total
\begin{align}
    I_2 &= - \lambda_0^\top (r_3 - p_3) + \chi_3 + \int_0^1 \frac{1}{2} J_2^\top \Lambda J_2 - J_2^\top (\chi_1 + t \chi_2) + \Gamma^\top  q_2 \d t, \\ \notag
    &= - \lambda_0^\top (r_3 - p_3) + \chi_3 + \int_0^1 \frac{1}{2} (t\Gamma + c)^\top L (t\Gamma + c) - \frac{1}{2} (\chi_1 + t \chi_2)^\top \Lambda^{-1} (\chi_1 + t \chi_2)\d t + \Gamma^\top  (p_2/2 + r_2/2 + s/6) \\ \notag
    &= - \lambda_0^\top (r_3 - p_3) + \chi_3 + 
    \frac{1}{2} c^\top L c + \frac{1}{2} c^\top L \Gamma + \frac{1}{6} \Gamma^\top L \Gamma   - \frac{1}{2} \chi_1^\top \Lambda^{-1} \chi_1 \\ \notag
    &\qquad - \frac{1}{2} \chi_1^\top \Lambda^{-1} \chi_2 - \frac{1}{6} \chi_2^\top \Lambda^{-1} \chi_2
    + \Gamma^\top  (p_2/2 + r_2/2 + s/6).
\end{align}
\subsubsection{Combining orders}
The above perturbation problem applies to general perturbations, but to compute the curvature we are only interested in a specific perturbation. In particular, calling $pr = \bar{W}(\rho_t,\rho_{t-\Delta t})$, $rq = \bar{W}(\rho_t,\rho_{t+\Delta t})$, and $pq = \bar{W}(\rho_{t+\Delta t},\rho_{t-\Delta t})$, we have that the local Menger curvature is 
\begin{equation}
    \kappa(t) = \lim_{\Delta t \to 0} \frac{\sqrt{(pq + pr + rq) (-pq + pr + rq) (pq - pr + rq) (pq + pr - rq)}}{pq \cdot pr \cdot rq},
\end{equation}
and moreover, we will show that in the limit 
\begin{align}
    rq &= \epsilon \sqrt{I_0^f} + \frac{\epsilon^2}{2} \frac{I_1^f}{\sqrt{I_0^f}} + \epsilon^3 \left[ \frac{1}{2} \frac{I_2^f}{\sqrt{I_0^f}} - \frac{1}{8} \frac{(I_1^f)^2}{(I_0^f)^{3/2}} \right] + \dots, \\
    pr &= \epsilon \sqrt{I_0^f} - \frac{\epsilon^2}{2} \frac{I_1^f}{\sqrt{I_0}} + \epsilon^2 \left[ \frac{1}{2} \frac{I_2^f}{\sqrt{I_0^f}} - \frac{1}{8} \frac{(I_1^f)^2}{(I_0^f)^{3/2}} \right] + \dots, \\
    pq &= 2\epsilon \sqrt{I_0^f}  +   \frac{\epsilon^3}{4} \frac{I_2^c}{\sqrt{I_0^f}}  + \dots,
\end{align}
and hence
\begin{equation}
    \kappa = 4 (I_0^f)^{-3/4} \sqrt{  \frac{I_2^f  - I_2^c/4}{\sqrt{I_0^f}} - \frac{1}{4} \frac{(I_1^f)^2}{(I_0^f)^{3/2}}},
\end{equation}
where we will now define the quantities $I_0^f$, $I_1^f$, $I_2^f$, $I_2^b$, and $I_2^c$.

To begin, consider the forward derivative, 
\begin{equation}
    \bar{W}(\rho_t,\rho_{t+\Delta t})^2 = \epsilon^2 I_0^f + \epsilon^3 I_1^f + \epsilon^4 I_2^f, \quad
    p = \rho_t + \Delta t \dot{\rho}_t + \frac{1}{2} \Delta t^2 \ddot{\rho}_t, \quad q = \rho_t,
\end{equation}
as well as the reverse derivative
\begin{equation}
    \bar{W}(\rho_t,\rho_{t-\Delta t})^2 = \epsilon^2 I_0^b + \epsilon^3 I_1^b + \epsilon^4 I_2^b,\quad
    p = \rho_t - \Delta t \dot{\rho}_t + \frac{1}{2} \Delta t^2 \ddot{\rho}_t, \quad q = \rho_t.
\end{equation}
and the centered difference
\begin{equation}
    \bar{W}(\rho_{t-\Delta t},\rho_{t+\Delta t})^2 = \epsilon^2 I_0^c + \epsilon^3 I_1^c + \epsilon^4 I_2^c, \quad
    p = \rho_t + \Delta t \dot{\rho}_t + \frac{1}{2} \Delta t^2 \ddot{\rho}_t, \quad q = \rho_t - \Delta t \dot{\rho}_t + \frac{1}{2} \Delta t^2 \ddot{\rho}_t.
\end{equation}
We have that $r_1^f = r_1^r = 0$, and $p_1^f = -p_1^r$, as well as $\Lambda^f = \Lambda_r = \Lambda^c$, so that $\Lambda$ and hence $L$ does not change. Therefore
\begin{align}
    I_0^f &=  \lambda_0^{f\top} p_1^f, \quad \text{where}\ L\lambda_0^f = p_1^f \\ \notag
    I_0^b &=  -\lambda_0^{b\top} p_1^f, \quad \text{where}\ L\lambda_0^b = -p_1^f \\ \notag
    I_0^c &=  4\lambda_0^{c\top} p_1^f, \quad \text{where}\ L\lambda_0^c = 2p_1^f,
\end{align}
so that $I_0^f = I_0^b = I_0^c/4$, and $J_1^f = -J_1^b = J_1^c/2$.

Next, at first order
\begin{align}
    I_1^f &=    p_2^\top \lambda_0^f -\sum_{e = (u,v)} \frac{J_1^f(e)^2}{4} \left( \frac{p_1(u)}{p_0(u)^2} + \frac{ p_1(v)}{p_0(v)^2} \right) \\ \notag
    I_1^b &=    p_2^\top \lambda_0^b -\sum_{e = (u,v)} \frac{J_1^b(e)^2}{4} \left( \frac{-p_1(u)}{p_0(u)^2} + \frac{ -p_1(v)}{p_0(v)^2} \right) \\ \notag
    I_1^c &=    0,
\end{align}
showing that $I_1^f = -I_1^b$. 

At second order, the contribution to $I_2$ from the term $-\lambda_0^\top (r_3 - p_3)$ satisfies $\lambda_0^{f\top} p_3 = \lambda_0^{b\top} (-p_3) = \frac{1}{4} \lambda_0^{c\top} (2 p_3) $, so that in the combination $I_2^f + I_2^b - I_2^c/2$, they cancel.

Next, note that as $\chi_1^f = \chi_1^b$, $\chi_2^f = \chi_2^b$, $\chi_2^f = \chi_2^b$, and $\Gamma^f = \Gamma^b$, we therefore have $s^f = s^b$, and so $c^f = c^b$. Therefore $I_2^f = I_2^b$. There is no straightforward way to relate $I_2^c$ to $I_2^f$ however, so in total three linear systems must be solved to find the curvature. This is still significantly cheaper than solving a second order conic, or even a minimum cost flow problem.
\\
\subsection{Local relation between \texorpdfstring{$\bar{W}$}{W} and \texorpdfstring{$d_{Diff}$}{dDiff}}
The diffusion distance between $p = p_0 + \epsilon p_1 + \epsilon^2 p_2 + \dots$, and $r = p_0 + \epsilon r_1 + \epsilon^2 r_2 + \dots$, is given by $d_{Diff}(p,q) = || D L^\dagger (p-r) ||$, with $q = t r + (1-t) p$ being one of many possible paths between them that is a geodesic under the $d_{Diff}$ distance. Under the metric $\bar{W}$, the geodesic path is given by
\begin{align}
    q =& p_0 + \epsilon \left( t r_1 + (1-t)p_1 \right) + \epsilon^2 \left( t r_2 + (1-t)p_2 + t(1-t) s \right) + O(\epsilon^3).
\end{align}
For this to be a geodesic path under $d_{Diff}$ as well, we would need that 
\begin{equation}
    || D L^\dagger (p-r) ||_1 = \sum_{i=1}^n d_{Diff}(q_{t_i},q_{t_{i-1}}),
\end{equation}
for all $0=t_0 < \cdots < t_n = 1$. However, we have that
\begin{equation}
d_{Diff}(q_{t_i},q_{t_{i-1}}) = (t_i - t_{i-1})|| D L^\dagger \left[ (p-r)  + \epsilon^2 (1- t_i - t_{i-1})s + O(\epsilon^3)\right]||_1.
\end{equation}
Taylor expanding for a single component, one has $|v + \epsilon^2 u + O(\epsilon^3)| = |v| + sign(v) \epsilon^2 u + O(\epsilon^3)$, and so we have that 
\begin{align}
    \sum_{i=1}^n d_{Diff}(q_{t_i},q_{t_{i-1}}) =& \sum_{i=1}^n (t_i - t_{i-1}) d_{Diff}(p,q) + \epsilon^2 sign(D L^\dagger (p-q)) \cdot D L^\dagger s \left[ t_i(1- t_i) - t_{i-1}(1-t_{i-1}) \right], \\\notag
    =& d_{Diff}(p,r) + O(\epsilon^3).
\end{align}
Thus, at least to order $\epsilon^3$, the unique geodesic under $\bar{W}$ is still a geodesic under $d_{Diff}$. Thus our curvature is truly a measure of how far the observed trajectory is away from the dissipation minimizing geodesic of $d_{Diff}$.

\subsection{Fitting an empirical curve with kernel density estimation}
Given a curve in the space of distributions on the flip graph, $p(t)$, we now have a way to compute the Menger curvature at a point, as a function of the first and second derivatives of the curve, $\kappa(t) = F(p(t),\dot{p}(t),\ddot{p}(t))$. However, we do not have access to the distribution, only empirical samples. We therefore wish to fit a smooth curve through these empirical samples, which we can then differentiate to compute the curvature. In the regime where samples are taken relatively far apart, but each sample is a good approximation to the true distribution, a spline can be fit through the samples~\cite{Clancy2021}. However, in the regime where the distributions may be undersampled, we do not wish to fit a curve through each sample. Instead, given samples $p_1,\dots,p_N$, at times $t_1,\dots,t_N$, we take a kernel density estimation approach and say
\begin{align}
    p(t) &= \sum_{i=1}^N a_i(t) p_i \\ \notag
     \text{with } a_i(t) &= \frac{\exp \left[ - \frac{(t - t_i)^2}{\sigma^2} \right]}{\sum_{j=1}^N \exp \left[ - \frac{(t - t_j)^2}{\sigma^2} \right] },
\end{align}
so that $p(t)$ is an average of the empirical observations, weighted to primarily include samples within a range of $\sigma$ away from $t$, and normalized to ensure $p(t)$ remains a probability distribution. Using automatic differentiation~\cite{AutoDiff}, it is then straightforward to find $\dot{p}$ and $\ddot{p}$ at any given $t$.

\subsection{Numerical validation}
\begin{figure}\centering
\includegraphics{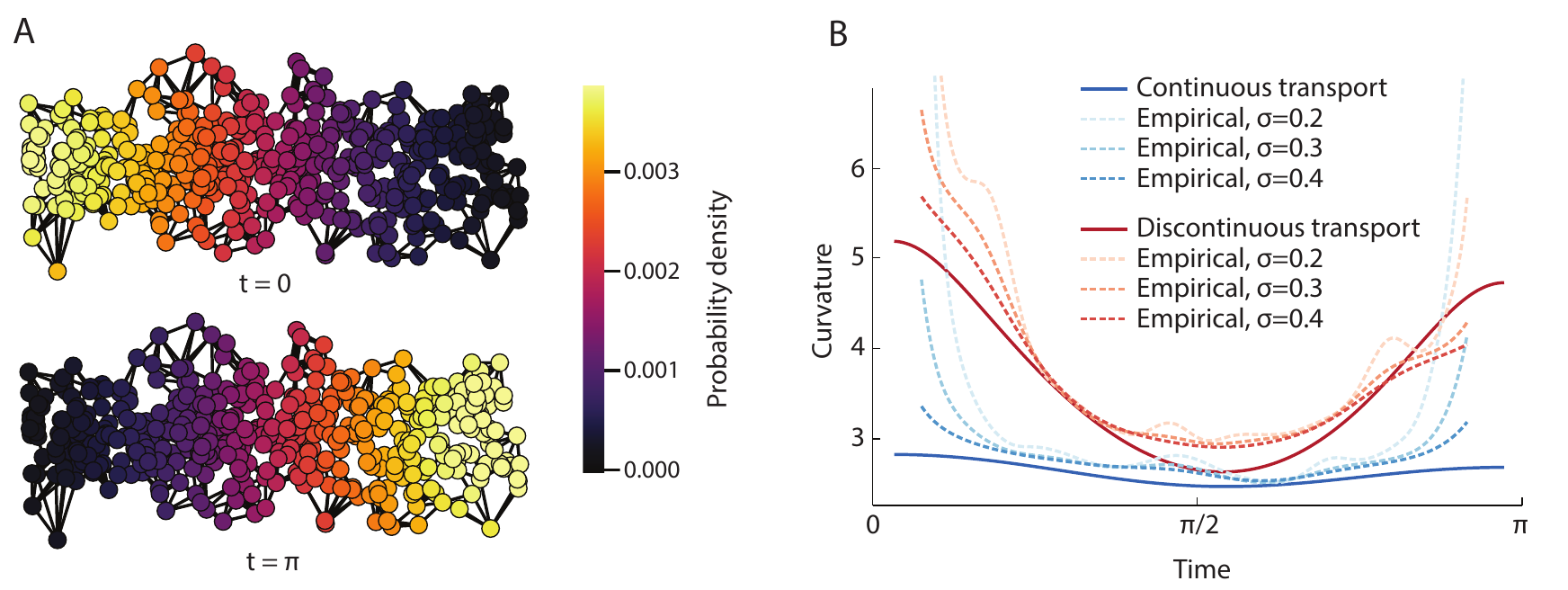}
\caption{\label{fig:MengerDemo}Computing the Menger curvature from empirical samples. (A) Random network with initial and final probability distributions at times $t=0$ and $t=\pi$. We will consider two different paths, $p_c$ and $p_d$ between these distributions. (B) Menger curvature of the continuous path, $p_c$ and the discontinuous path $p_d$, along with the empirical approximations from finite sampling.}
\end{figure}

In order to validate the curvature framework, we introduce here a model system. We create a network by sampling 500 points uniformly in $x\in [-0.5,0.5]$ and as $\mathcal{N}(0,0.1)$ in $y$, and connecting each point to their 7 nearest neighbors. We consider two different paths on this graph, 
\begin{align}
    N(x,y) =& \ e^{-3(x^2+y^2)} \\ \notag
    p_c(t,x,y) \propto& \ N(x-\cos t,y)\\ \notag
    p_d(t,x,y) \propto& \ (1+\cos t)N(x-1,y) + (1-\cos t)N(x+1,y),
\end{align}
where $p_c$ and $p_d$ are only defined for $x$ and $y$ where there is a vertex, and there is a proportionality factor ensuring that $p_c$ and $p_d$ always remain normalized. Intuitively, $p_c$ represents a continuous path, where probability density is transported from the left hand side of the network (Fig.~\ref{fig:MengerDemo}A) at $t=0$ to the right hand side at $t=\pi$ by continuously shifting it from left to right. In contrast, $p_d$ represents a discontinuous path where probability density instantaneously switches from the left side to the right side. Both distribution agree at initial and final time points, $p_c(0) = p_d(0)$, $p_c(\pi) = p_d(\pi)$.

Intuitively we would expect $p_c$ to have lower curvature than $p_d$, and we confirm this by computing the curvature exactly from the functions $p_c$ and $p_d$, Fig.~\ref{fig:MengerDemo}B. Moreover, we would expect the curvature of $p_c$ to be relatively constant in time, whereas we would expect the discontinuous path to have largest curvature at the start and end times, where probability density must appear in an area with previously very low probability, Fig.~\ref{fig:MengerDemo}B. Note that the reflection symmetry around $t = \pi/2$ is broken by the normalization and the random distribution of points. 

Now we draw finite samples from this distribution. At time intervals of $t=0.1$, we draw 8000 samples from each distribution, a notable undersampling. From this, we use kernel density estimation to construct a continuous curve and compute the Menger curvature. We approximately recover the curvature, Fig.~\ref{fig:MengerDemo}B, although taking too small of a $\sigma$ value leads to spurious oscillations, especially around the end points.
\nolinenumbers
%\linenumbers

\if 0
\subsection{Ventral furrow curvature}
We apply the curvature framework to the formation of the ventral furrow in \emph{Drosophila} embryogenesis. Nuclei of cells are identified with ? microscopy at time resolution of ? minutes  over the time in which the ventral furrow forms~\ref{fig:FurrowCurvature}A. Combining consecutive time points, and computing the pairwise distances between the subsequent distributions recovers the temporal ordering in the MDS embedding~\ref{fig:FurrowCurvature}B. The MDS embedding suggests that formation of the furrow (from around time 15 to 35) is inefficient as it loops back on itself with time 10 being very close to time 35. However, the curvature shows that this is an artifact of projecting into a low dimensional Euclidean space, and that during the furrow formation, locally, the path taken is  maximally efficient~Fig.~\ref{fig:FurrowCurvature}C.

\nolinenumbers
\begin{figure}\centering
\includegraphics{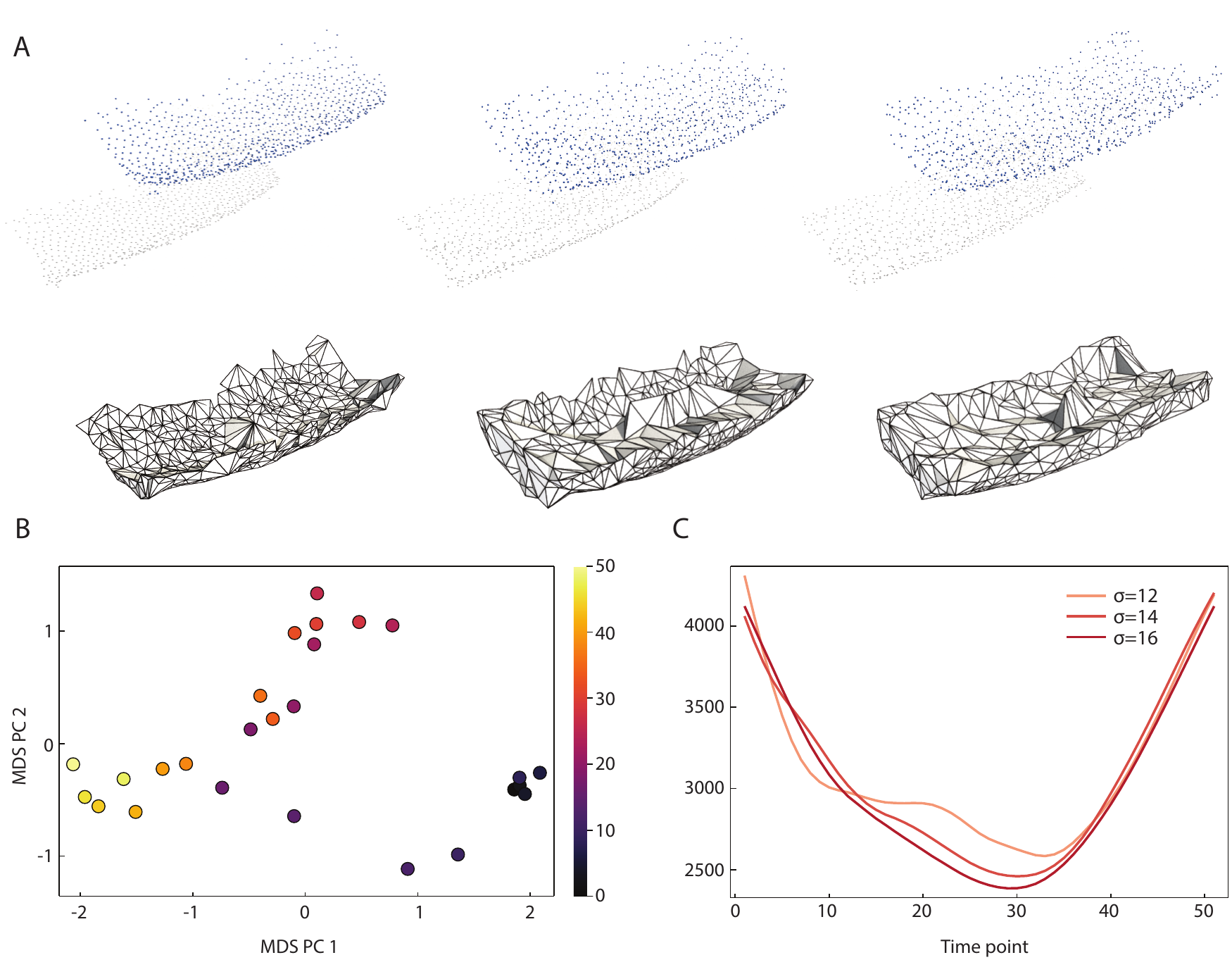}
\caption{\label{fig:FurrowCurvature}Computing the Menger curvature for the ventral furrow curvature. (A) Nuclei around the region of the ventral furrow shown at various times (top) with corresponding alpha complex (bottom). TODO: Scale bar, time points, and maybe change angle. (B) MDS embedding from distributions created by combining two consecutive time points, recovers the temporal ordering. (C) Curvature for various $\sigma$ values shown.}
\end{figure}
%\linenumbers
\fi

\subsection{Zebrafish curvature}
\begin{figure}
\centering
\includegraphics{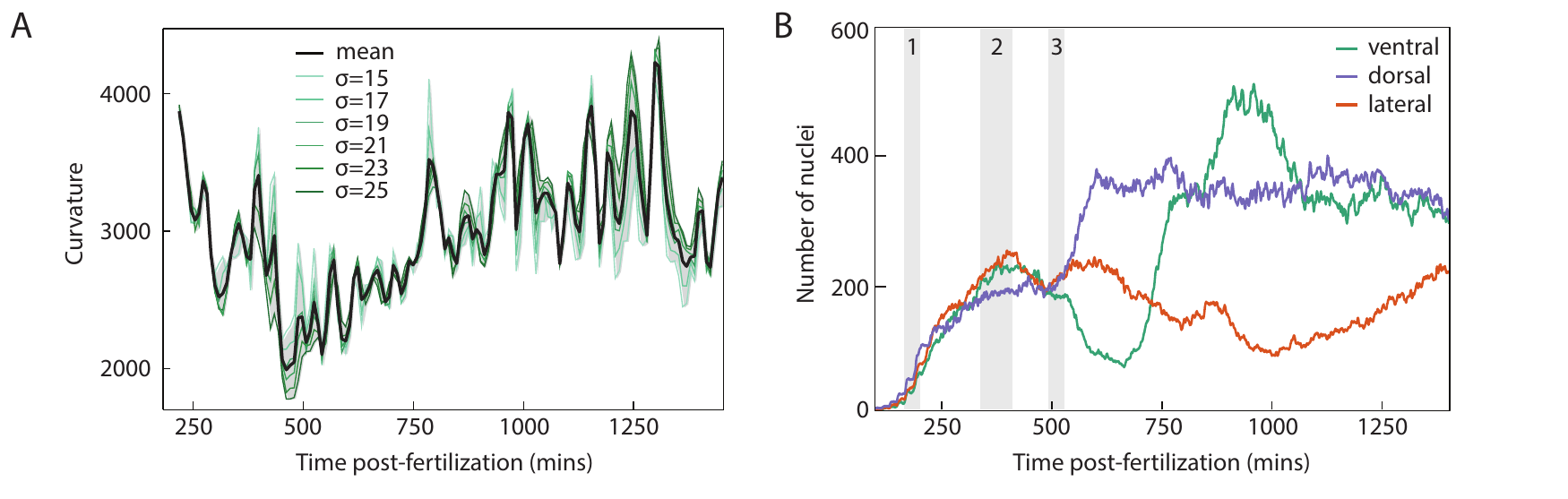}
\caption{\label{fig:ZebCurvature}Topological efficiency of developmental trajectory is revealed by a curvature computation for zebrafish embryogenesis. (A) To analyze the topological trajectory without using a low dimensional embedding, which will necessarily distort the data, we compute the curvature, a measure of how efficient the topological path is (higher curvature is less efficient).  To compute the curvature, kernel density estimation was used with different smoothing parameters $\sigma=15,\dots,25$, from which a mean value was computed (black). (B) Number of cell nuclei in dorsal ventral and lateral regions of the embryo, adapted from Ref.~\cite{Keller2008}. Three symmetry breaking events (gray regions) were identified by Ref.~\cite{Keller2008}, namely: (1) symmetry breaking in cell divisions, (2) symmetry breaking in cell density, (3) symmetry breaking in morphology. Following cell numbers in each region after these initial symmetry breaking events provides a partial way to identify developmental changes, albeit limited as it only tracks cell numbers and not cell rearrangement or topology.
}
\label{fig:Curvature}
\end{figure}

After computing the pairwise distance matrix between 90 time points of the zebrafish embryo, we recover a 1D manifold that is parameterized by time, Fig 2D. We find that development changes the topological structure, and that these developmental changes are more significant than finite sampling effects as minimal fluctuations around this path are present in the embedding. The path that this manifold takes is not a straight line in the embedding. Contrast that with the developmental trajectory of a 2D fly wing, which was found to take an optimal path from the initial distribution to the final one~\cite{Skinner2021}. To analyze where the path is least optimal, without the distorting effects of a low dimensional embedding, we compute the curvature. After starting at a large curvature, the curvature drops at around 500 m.p.f., corresponding to the straightest part of the MDS embedding, before increasing again, Fig.~\ref{fig:ZebCurvature}A. Spikes in curvature can correspond to abrupt changes the topological trajectory, and could correspond to abrupt developmental changes. To further investigate this, we can compare this to the nuclei counts in three domains, dorsal, ventral, and lateral, that were used in Ref.~\cite{Keller2008} to identify symmetry breaking events, Fig.~\ref{fig:ZebCurvature}B. The first event identified by Ref.~\cite{Keller2008}, is symmetry breaking in cell divisions, which happens early on when the curvature is high. The next corresponds to symmetry breaking in cell density which happens at a local maximum of curvature. After this, the curvature drops to it's lowest point as the total number of cells remains flat, with mostly rearranging rather than cell division occurring, before the final identified symmetry breaking event of the first morphological differences between regions. The next notable spike in curvature comes at around 800 m.p.f., where a rapid growth of cells in the ventral region stops, marking the end of some particular developmental phase and the beginning of another. Subsequent spikes in curvature become hard to associate with a particular developmental change, as identified by coarse grained nuclei counts, as by this point, many independent processes are occurring across the embryo.

%%%%%%%%%%%%%%%%%%%

\section{Bootstrapping distance calculations}\label{sec:Bootstrap}
Whilst the MDS embedding shows that different species of biofilm have different topological distributions, we would like to test statements of that form statistically. Consider the example of comparing \emph{E. coli} and \emph{V. cholerae}, where we have 15 \emph{E. coli} experiments and 15 \emph{V. cholerae} experiments. To compare the two species, we can combine all 15 \emph{E. coli} experiments into a single observed distribution and compute the distance with the combined 15 \emph{V. cholerae} experiments. How can we know whether the resulting distance is statistically significant? In particular, due to finite sampling effects, we would expect there to be a non-zero distance between even 15 combined \emph{E. coli} experiments and another independent set of 15 combined \emph{E. coli} experiments; the empirical distributions would not perfectly match even though there is no underlying structural difference. However, by comparing against a null distance, we are able to perform significance tests on the hypothesis that the distance between two species is non-zero.

\begin{figure}\centering
\includegraphics{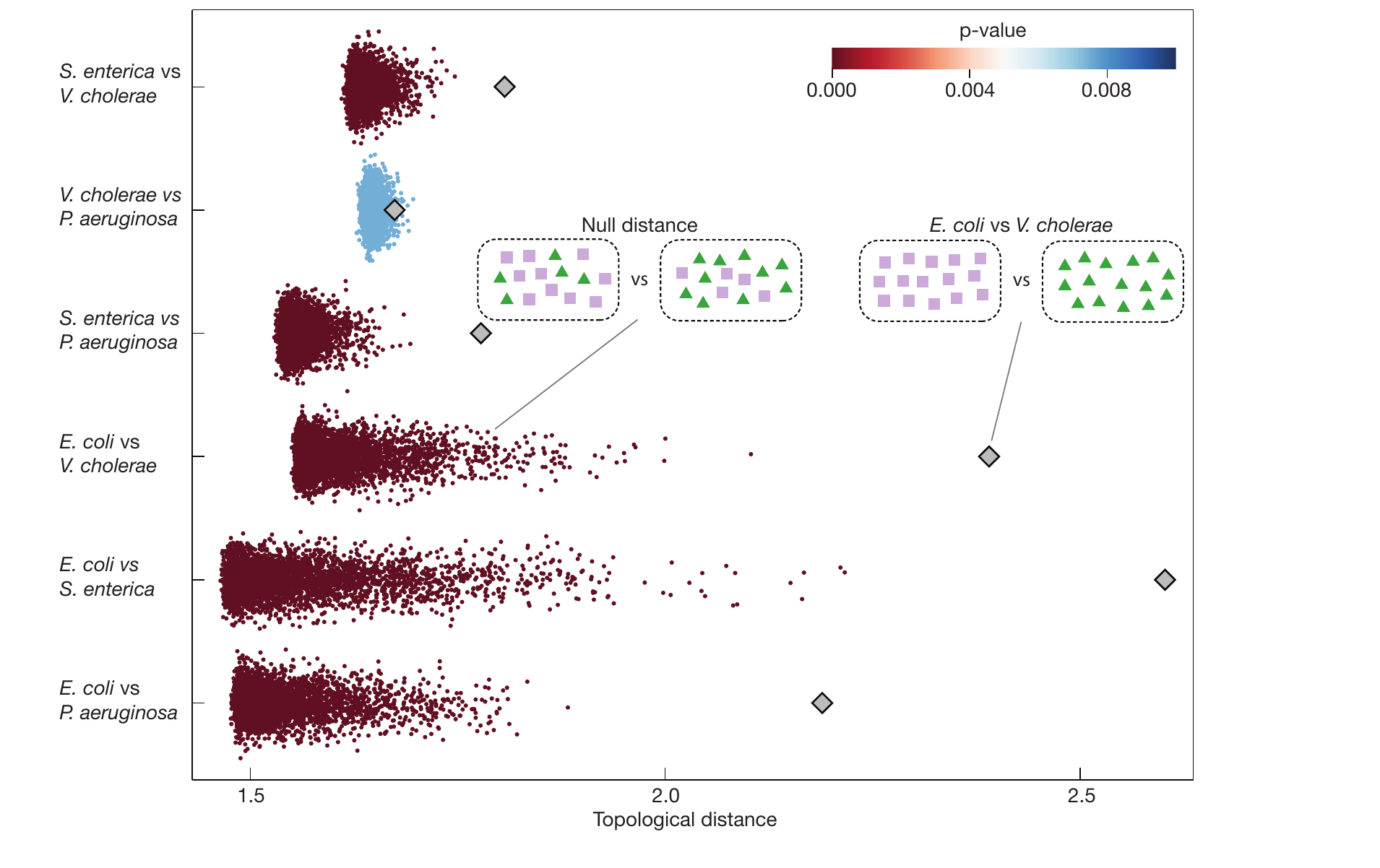}
\caption{\label{fig:BiofilmBootstrap}Topological distance between different bacterial species is statistically significant at $p<0.01$. For each pair of species the topological distance was computed between the combined distribution of 15 experiments from the first species and the combined distribution of 15 experiments from the second species (gray diamonds). This was compared to a null distance which was formed by randomly splitting the 30 combined experiments into groups of two, which was done 2000 times (points, colored by p-value). In all cases the distance between species is greater than $99\%$ of null distances. }
\end{figure}

Assume for a moment the null hypothesis, that there is no structural differences between \emph{E. coli} and \emph{V. cholerae}. In this case, there is nothing special about splitting the 30 experiments into 15 \emph{V. cholerae} and 15 \emph{E. coli} experiments, as they are all structurally identical. There are many other ways to divide these 30 experiments into two, for instance one group with 10 \emph{E. coli} and 5 \emph{V. cholerae} experiments, and another group with 5 \emph{E. coli} and 10 \emph{V. cholerae} experiments. In fact, there are ${30 \choose 15}/2 = 77558760$ ways of dividing the 30 experiments into two equally sized groups. If the null hypothesis holds, there is nothing special about the distance between 15 \emph{V. cholerae} and 15 \emph{E. coli} experiments, as compared to distances resulting from all the other ways to split the experiments. However, when we compute the distances between 2000 randomly sampled different ways of splitting the 30 experiments, the \emph{V. cholerae} vs \emph{E. coli} split is larger than any of them, Fig.~\ref{fig:BiofilmBootstrap}.

We can repeat this for all match ups of different biofilm species, finding for all of them that every pairwise combination of species is statistically significantly different for at p-value $p < 0.01$, meaning that the pairwise distance when split by species is bigger than $99\%$ of random splittings. We therefore can conclude that every one of the 4 species considered has a topologically distinct structure from any of the others. Whilst shown to be topologically distinct, \emph{P. aeruginosa} and \emph{V. cholerae} form biofilms that are more similar structurally than any other pair of species, with both having similar aspect ratio cells.

We can also apply this analysis to the different regions of the juvenile zebrafish brain, as was done in  Fig.~2 of the main text. Nuclei within the brains were categorized through a semi-automated histological approach into 9 major brain regions~\cite{Cheng2019}, and as a point of comparison, we also divided the nuclei into 9 regions along the principal head--tail axis, with each region containing an approximately equal number of points. With only 5 experiments available, we can not say that all major brain regions are statistically different, but at $p<0.05$ we can say that 32 (out of 36) pairwise comparisons are significantly different. In comparison, for the regions created by partitioning along the major axis, only 8 pairwise comparisions are statistically significant. Further, all of these 8 comparisons involve region 1, which is effectively the olfactory epithelium and telencephalon regions combined, whereas all other partitioned regions are a mix of several different major brain regions.
\nolinenumbers
%\linenumbers

\nolinenumbers
\begin{figure}\centering
\includegraphics{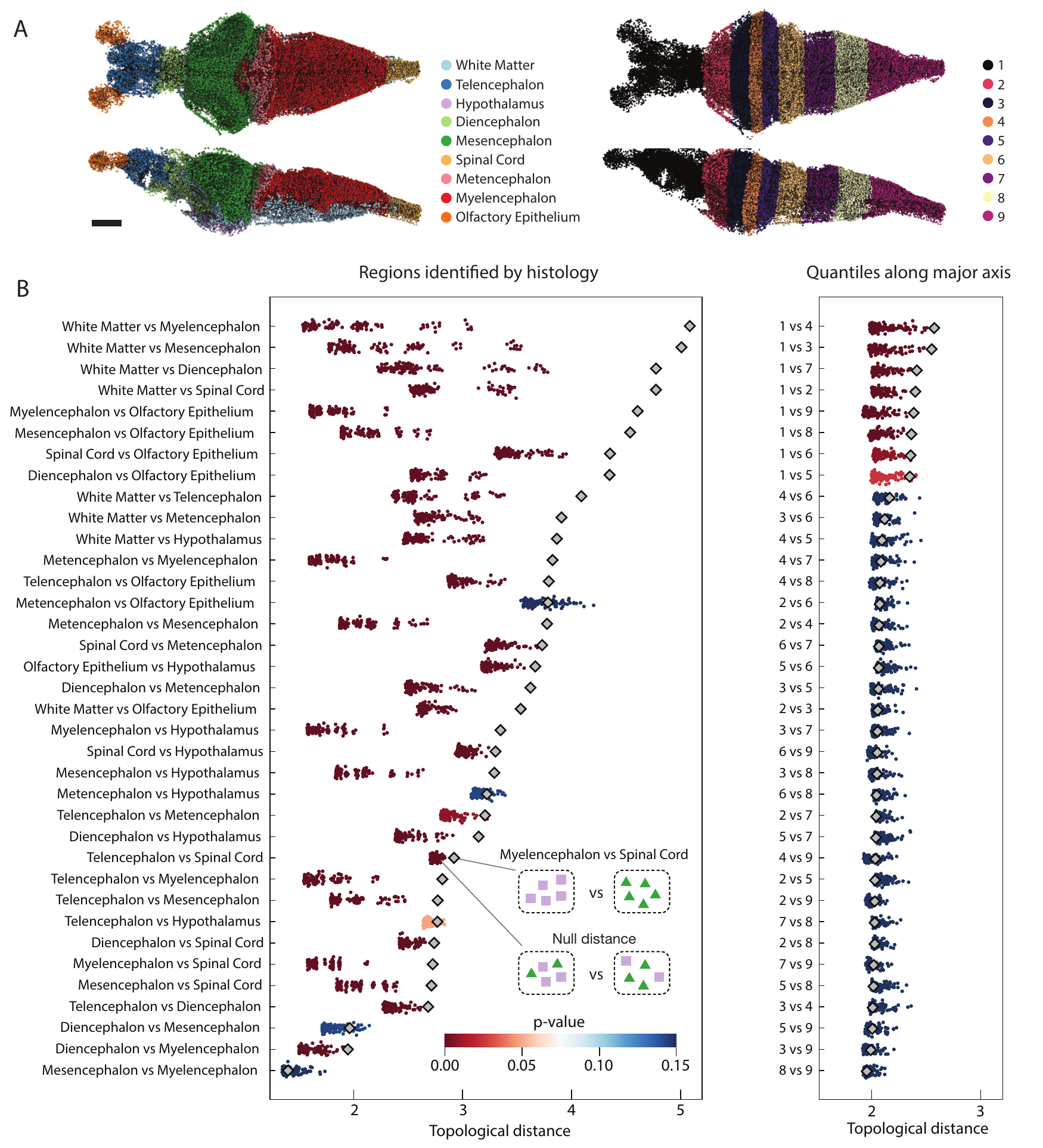}
\caption{\label{fig:ChengBootstrap}
Regions of the zebrafish brain identified by histological imaging differ in a statistically significant way, in contrast to regions created by partitioning along the major axis. (A) Zebrafish brain nuclei from Ref.~\cite{Cheng2019} colored by the 9 major brain regions (left) and by partitioning into 9 regions with equal nuclei along the major axis. Scale bar is 100$\mu$m. (B) For every pair of zebrafish brain regions, the topological distance was computed between the combined distribution of the first region taken from 5 experiments and the combined distribution of the second region taken over 5 experiments (gray diamonds). This was compared to a null distance which was formed by randomly splitting the 10 regions (2 from each experiment) into two groups of 5, which was done all 126 possible times (points, colored by p-value). This was done for the 9 major brain regions identified by Ref.~\cite{Cheng2019} with histological imaging analysis (left), and for 9 regions created by partitioning along the major axis (right). With only 5 experiments, not all pairs of regions can be shown to differ in a statistically significant way. However, for the major brain regions, 32 pairs (out of 36) differ at $p<0.01$, whereas only 8 pairs do for the major axis partitioning.
}
\end{figure}
\section{Details of datasets used}

\begin{figure}\centering
\includegraphics[width=0.9\textwidth]{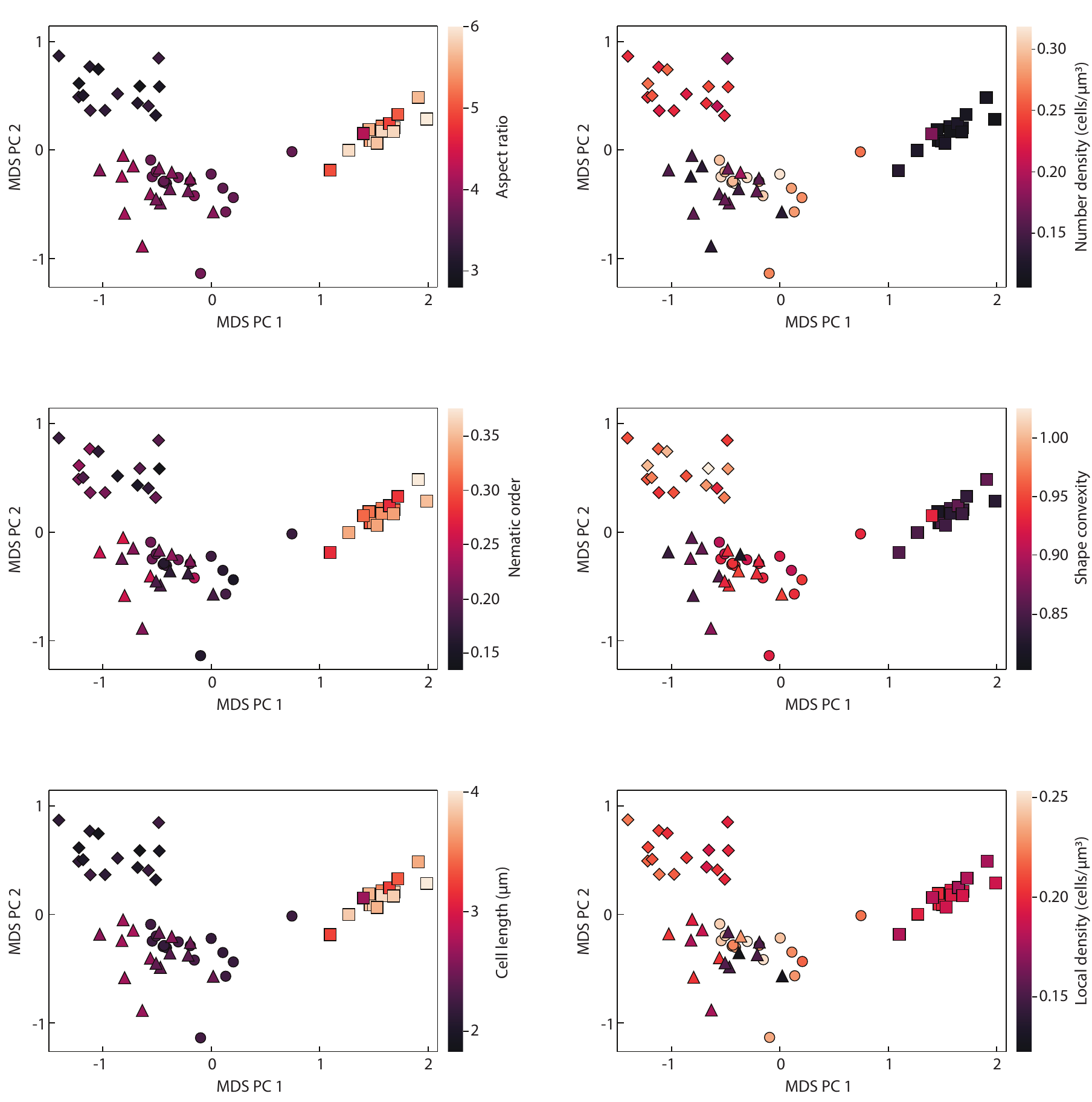}
\caption{\label{fig:BiofilmMDS}MDS embedding of biofilm data colored by single cell properties finds that aspect ratio explains the data manifold. For every biofilm experiment, the cell aspect ratio, cell length, and shape convexity were calculated for each cell, and the mean was taken to create an average value for the experiment. The remaining parameters were computed by drawing a ball for every cell of radius $2\ \mu$m and computing the nematic order parameter, the local number density (cells per volume), and the density (volume occupied by cells), within that ball. These were again averaged to create a value for each experiment. Coloring by aspect ratio and the related quantities of cell length and nematic order parameter, provides a consistent coloring along the data manifold (left). Coloring by other quantities does not (right). This suggests that differing cell aspect ratio is the principal reason for topological differences between colonies.}
\end{figure}

\textbf{Biofilms.} Bacterial biofilms are multicellular communities of cells that grow on surfaces and are held together by an extracellular matrix~\cite{HallStoodley2004}, and are among the most abundant forms of microbial life on earth~\cite{Flemming2019}. We make use of the segmented images of bacterial biofilms from Ref.~\cite{BiofilmStructure}, with 15 biofilms imaged from the four species, \emph{E. coli, V. cholerae, S. enterica, P. aeruginosa}. The segmented colony contains the size, position, and orientation of every cell within the colony~\cite{BiofilmStructure}, here we use only the cell centroid position. The biofilms are imaged whilst growing, and for each experiments we have a number of time points. We use only the images when the biofilm contains between 1500 and 3500 cells, typically resulting in around 4 images per biofilm. We combine all the motifs from each image of the same biofilm into one distribution. After calculating the distance $d_\text{Diff}(A,B)$, for all pairs of biofilms $A,B$, we embed this distance matrix in 2D using multi-dimensional scaling, and can color according to single cell properties, Fig.~\ref{fig:BiofilmMDS}. We see that cell aspect ratio, and related quantities like cell length, color the principal component of the manifold on which the data lie.

%We also analyze the distribution of motif sizes, where the size of a motif is defined by the number of tetrahedrons it contains.

\nolinenumbers
%\linenumbers

\textbf{Zebrafish brain.} Juvenile zebrafish brains were imaged by micro-CT tomography and the spatial position of all nuclei were segmented in Ref.~\cite{Cheng2019}. In total there are 5 experiments available, each containing around 80,000 nuclei.  In addition to nuclei segmentation, Ref.~\cite{Cheng2019}, applied a semi-automated histological approach to assign each nuclei to one of 9 major brain regions. Motifs were computed prior to assigning the into major brain regions, so if the central vertex of a motif lies in one region, that motif is assigned to that region even if it contains vertices in different regions.

\begin{figure*}%[\sidecaptionrelwidth][t]
\centering
\includegraphics{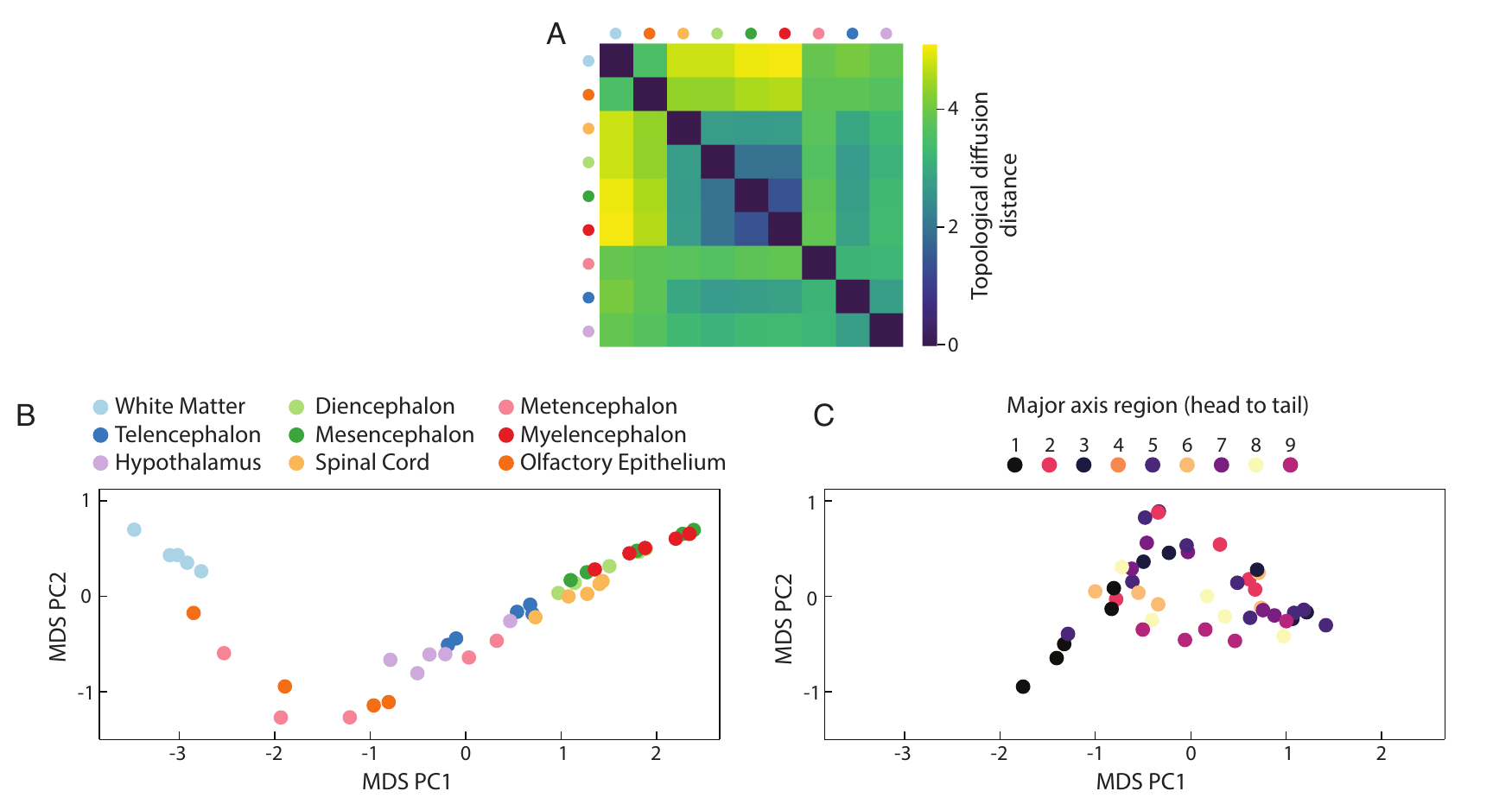}
\caption{Histologically identfied regions of the juvenile zebrafish brain show systematic topological differences. (A) Combining all experiments to compute a distance between regions, we see that some regions are far more topologically different than others. For instance, the white matter region is very different than the Myelencephalon region ($p < 0.01$), but the Myelencephalon region is topologically similar to Mesencephalon region (no statistically significant difference). (B) Embedding all regions of the brain for each experiment separately reveals consistent differences between brain regions across experiments, as well as a broadly one-dimensional manifold along which regions vary. (C) In contrast, dividing into 9 equally sized regions based on the distance along the major axis reveals no significant difference between regions.}
\label{fig:Fish}
\end{figure*}

For the 5 experiments available, and for every major brain region, we take a combined motif distribution using the nuclei assigned to that region across all experiments. Using these distributions, we compute the pairwise distance matrix between every brain region, Fig.~\ref{fig:Fish}A. We see that, for example, the distance between white matter and Myelencephalon is over 10 times the difference between Mesencephalon and Myelencephalon. Indeed, from Fig.~\ref{fig:ChengBootstrap}, we see that the distance between white matter and Myelencephalon is statistically significant at $p<0.01$, whereas the distance between the Mesencephalon and Myelencephalon is not statistically significant at this (limited) level of data. 

In contrast, taking the 9 regions along the major axis, across experiments, the embedding does not show systematic variation across regions, with variation between experiments playing as important of a role, Fig.~\ref{fig:Fish}C.
%%%%%%%%%%%%%%%%%%%

%%%%%%%%%%%%%%%%%%

\begin{figure}\centering
\includegraphics{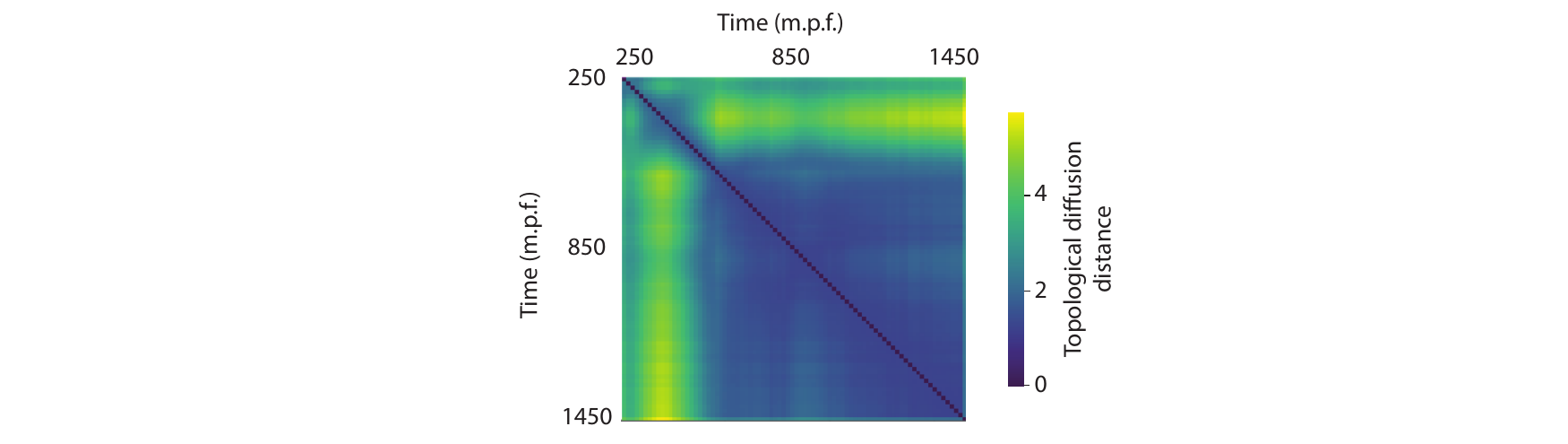}
\caption{\label{fig:KellerDistance} Topological distance matrix for the zebrafish embryo showing pairwise distances between 90 time points. The resulting MDS embedding of this distance matrix is shown in Fig 2D.}
\end{figure}

\textbf{Embryo development.}
A developing zebrafish was imaged using lightsheet microscopy and the position of all nuclei were determined from around 100 to 1500 minutes post fertilization by Ref.~\cite{Keller2008}. In total, 900 time points were imaged at regular intervals separated by 90s. For the MDS embedding, the 900 time points were split into 90 regions, each containing 10 subsequent time points, and the pairwise distance was calculated between these regions, as shown in Fig.~\ref{fig:KellerDistance}. For the combined embedding, only 6 of these 90 regions were used, to illustrate different developmental stages.

\nolinenumbers
%\linenumbers

The same group imaged a developing \emph{D. melanogaster} embryo with lightsheet microscopy and again collected the position of all nuclei from 120 to 690 minutes post fertilization by Ref.~\cite{Keller2010}. In total, 191 time points were imaged at regular intervals separated by 180s. For the combined embedding, only 7 time points were used. All times for both the zebrafish and \emph{D. melanogaster} are measured in minutes post-fertilization (m.p.f.).

We also made use of two additional experiments of embryonic development, the ascidian \emph{P. mammillata} and the worm \emph{C. elegans}. Detailed images of the developing ascidian were collected by Ref.~\cite{AscidianScience} across 6 experiments. We used all images containing over 250 cells, and took the cell barycenters as input points for the Delaunay. For \emph{C. elegans}, 46 experiments were performed and nuclei positions obtained by Ref.~\cite{Cao2020}. For each of these experiments, we an image which occurred at around the 350 cell stage.

\textbf{Human Cancer Organoid.} The mechanical properties and physical arrangement of human cancer cells determine their ability to invade surrounding tissues~\cite{GuoOrganoid}. In total, 18 human cancer organoids were imaged by Ref.~\cite{GuoOrganoid} at the 21 day stage, comprised of around 100-400 cells, with the nuclei detected. We used the spatial location of the nuclei for the combined embedding.

\textbf{Random packings.} Random packings of spheres and ellipsoids were generated using the event driven packing code from Ref.~\cite{Donev2005}. This created a jammed periodic packing of 10,000 particles, where the centroids of the particles were used for subsequent analysis. The periodicity allows us to compute the motif for every point without boundary effects, since every point can be considered to be in the bulk. Simulations were performed for spheres, $1:1:4$, $1:4:4$, and $1:2:3$ aspect ratio ellipsoids. Simulations were also performed for a polydispere mixture of spheres with radii in ratio $2:3$, and with equal numbers of each.

\textbf{Diffusion limited aggregation.} In order to model the process by which particles combine in order to form an aggregation, for instance in dust or soot, a mathematical model known as diffusion limited aggregation was introduced and has been widely studied~\cite{DLAWitten}. Simulations of diffusion limited aggregation were performed with code from Ref.~\cite{DLA_code} with the default parameters.

\textbf{Glassy material.} Under certain conditions, liquids can be cooled down to a solid-like state without forming local crystaline order. Understanding the nature of this transition, as well as the properties of such glassy systems is a central challenge in physics~\cite{GoogleGlassy}.
Simulations of a glassy material specifically a 80:20 Kob–Andersen-type Lennard–Jones mixture were performed by Ref.~\cite{GoogleGlassy} with 4096 particles in a periodic box. We used the final time point of simulations performed at temperature $T=0.44$ , well into the glassy phase.

\textbf{Star positions.}
The positions of the nearest $110,000$ stars to earth were taken from the HYG star database~\cite{HYGDatabase}, which collates three previous databases, Refs~\cite{1991GLIESE, 1995YALE, 1997HIPPARCOS}. Within this database, only the stars with known position, and not just angle, were used. In the computation of the motif distribution, motifs at the edge of this dataset were removed, as this is not a physical boundary.

\subsection{Combined embedding and residual variance}
\begin{figure}\centering
\includegraphics{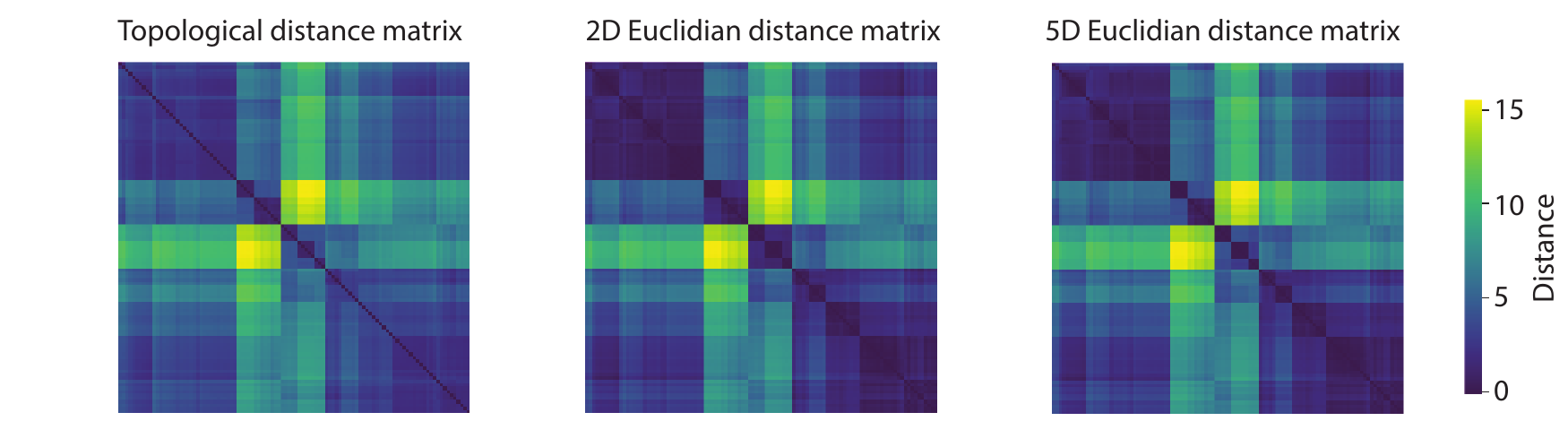}
\caption{\label{fig:DistanceDistortion} Topological distance matrix is not distorted by embedding into Euclidean space. The topological distance matrix for living and non-living systems (left) was embedded into Euclidean space with MDS. Computing the Euclidean distance matrix of the resulting MDS embedding, shows that in 2D (middle), the original distance structure is largely preserved. The main difference is that similar regions are forced closer together in the low dimensional embedding. Embedding into a higher dimensional Euclidean space improves the fidelity of the distance matrix (right for 5D embedding), for instance compare the center region of distance matrices).}
\end{figure}

When embedding a (non-Euclidean) distance matrix into a lower dimensional Euclidean space, we must ensure that the  lower dimensional representation is not significantly distorting the true distance matrix. To do so, we can compare the true topological distance matrix with the distance matrix of the embedding, calculated by measuring the pairwise Euclidean distances of the embedding. We see that the Euclidean distances of the embedding remain faithful to the true distances, Fig.~\ref{fig:DistanceDistortion}. The primary distortion is that some similar systems are forced closer together in the embedding than they actually are in distance, which can be improved by taking a higher dimensional embedding, Fig.~\ref{fig:DistanceDistortion}.

\nolinenumbers
%\linenumbers

\subsection{Comparing living and non-living systems}
From the combined embedding in Fig. 3, we see that the region formed by taking the convex hull of all  living systems only contains one non-living system in a 2D embedding, and contains no non-living systems in a 3D embedding.

\nolinenumbers
\begin{figure}\centering
\includegraphics{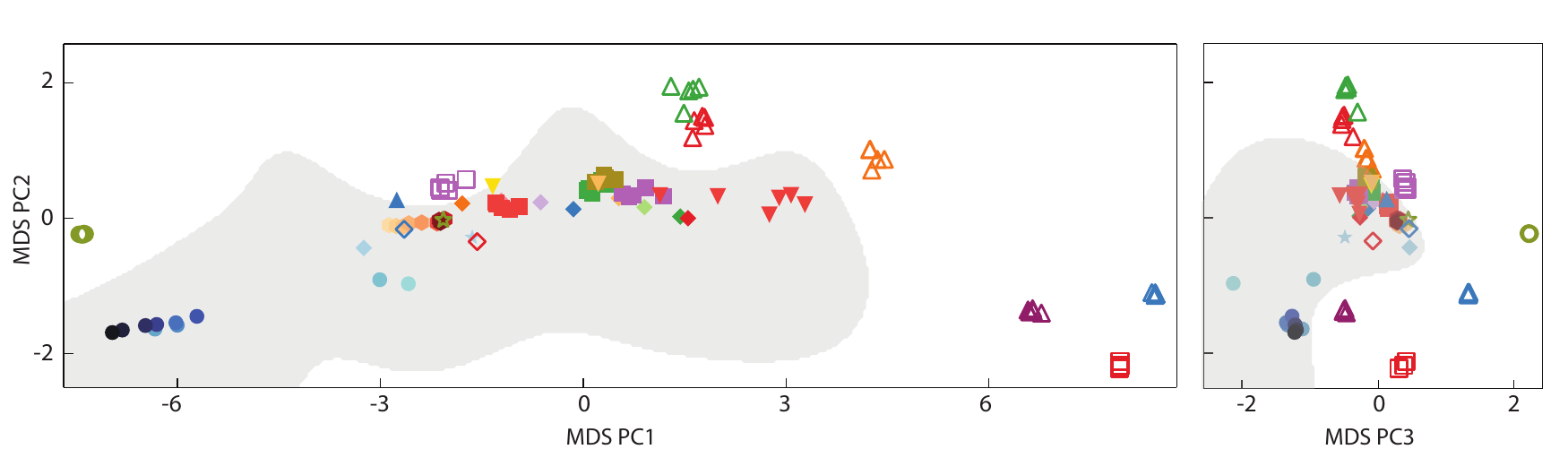}
\caption{\label{fig:SVM} Support vector machine with a non-linear kernel trained on the combined MDS embedding can classify living and non-living materials. From the first two components of the MDS embedding (left) a support vector machine with a single exponential radial basis function non-linear kernel was trained to classify living (closed symbols) and non-living (open symbols) systems. The region it assigns to living systems is shown in gray, and gets the classification correct with the exceptions of the galaxy,  polyurethane foam, and fluid foam data sets. Repeating this procedure using components 2 and 3 of the MDS embedding (right) the support vector machine is similarly able to correctly classify points with the exception of the galaxy,  polyurethane foam, and fluid foam data sets and one of the irregular ellipsoid packings (for full symbol legend see Fig. 3).}
\end{figure}
%\linenumbers

Additionally, we can test whether there exists a linear hyperplane separating living and non-living systems by training a support vector machine~\cite{SVM_tutorial} on the embedding data. We find that in a 7 dimensional MDS embedding, a linear hyperplane exists which correctly classifies all points except the star database. Moreover, using a support vector machine with a non-linear kernel (exponential radial basis function), we can train a function that correctly classifies all but the star database and organoid data from a 2D MDS embedding, or all but the star database and two irregular ellipsoid packings from MDS components 2 and 3, Fig.~\ref{fig:SVM}.

\subsection{Motif distribution analysis}
\begin{figure}\centering
\includegraphics{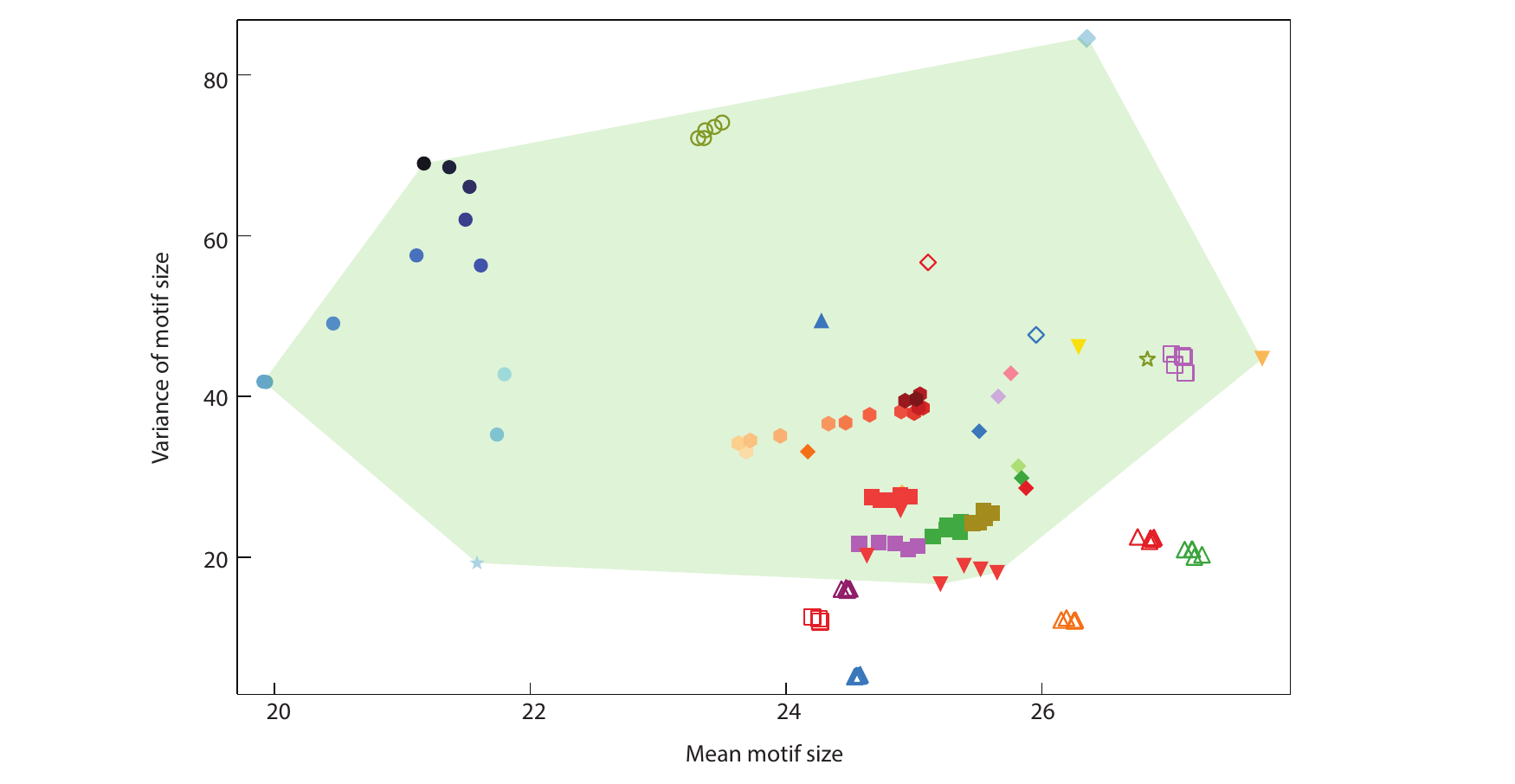}
\caption{\label{fig:MomentEmbedding} Plotting mean motif size against variance of motif size reveals systematic differences across systems. Symbols are the same as main text Fig. 3, convex hull of living systems is shown in green. Multiple realizations of the same process, such as Poisson-Voronoi (purple squares) show that these statistics are not significantly affected by finite sampling effects. }
\end{figure}

In addition to the combined embedding and distance computation, it is of interest to investigate how the motif distributions vary. It is hard to visualize a distribution over the total space of motifs, instead we can compute simpler statistics of the distributions and study these. For instance, we can compute the mean motif size, which counts on average the number of tetrahedrons in a motif, as well as computing the variance of this quantity. We find that these systematically vary across different systems, Fig.~\ref{fig:MomentEmbedding}. Whilst some of the features of the combined embedding are also present in this moment embedding, the convex hull of living systems contains a number of non-living systems, Fig.~\ref{fig:MomentEmbedding}. Moreover, systems that are only somewhat different in the mean and variance of motif sizes are in actuality very different according to the TDD distance, for instance the polydisperse packing and the biofilms, Fig.~\ref{fig:MomentEmbedding}.
\nolinenumbers
%\linenumbers

\clearpage
\section{Tables}

\begin{footnotesize}
\begin{longtable}{p{3.2cm} p{1.cm}@{\hskip 0.6cm} p{6.5cm}@{\hskip 0.6cm} p{3.1cm}@{\hskip 0.3cm} p{2.2cm} }
\caption{\label{tab:Comb}Parameter summary for combined topological embedding}\\
Region & Num. samples & Num. points & Boundary protocol & Data source \\ [1ex] 
\hline\hline & \\[-2ex]
%%%%%%%%%%%%%%%%%%%%%%%%%%%%%%%
Bacterial biofilms & & & & \\ 
\emph{V. cholerae} & 5 &  3 combined exp. with $\sim 3$ time points each with $\sim2000$ cells  &  $\alpha = 4\mu m$ & Ref.~\cite{BiofilmStructure} \\ 
\emph{E. coli} & 5 &  \ditto & \ditto & \ditto  \\ 
\emph{S. enterica} & 5 &  \ditto & \ditto & \ditto \\ 
\emph{P. aeruginosa} & 5 &  \ditto & \ditto & \ditto  \\ 
\hline& \\[-2ex]
%%%%%%%%%%%%%%%%%%%%%%%%%%%%%%%
Zebrafish brain region & & & & \\ 
Olfactory Epithelium & 1 & 5 combined exp. each with $\sim1000$ cells & $\alpha=30\mu$m & Ref.~\cite{Cheng2019} \\ 
Telencephalon & 1 &  5 combined exp. each with $\sim~4000$ cells & \ditto & \ditto  \\ 
Diencephalon & 1 & 5 combined exp. each with  $\sim~7000$ cells & \ditto & \ditto \\  
Hypothalamus & 1 & 5 combined exp. each with $\sim~2500$ cells & \ditto & \ditto  \\ 
Mesencephalon & 1 & 5 combined exp. each with $\sim~18000$ cells & \ditto & \ditto  \\  
Metencephalon & 1 & 5 combined exp. each with $\sim~1500$  cells& \ditto & \ditto \\ 
Myelencephalon & 1 & 5 combined exp. each with $\sim~30000$ cells & \ditto & \ditto  \\ 
White Matter & 1 & 5 combined exp. each with $\sim~5000$ cells & \ditto & \ditto \\ 
Spinal Cord & 1 & 5 combined exp. each with $\sim~1500$ cells & \ditto & \ditto  \\
\hline& \\[-2ex]
%%%%%%%%%%%%%%%%%%%%%%%%%%%%%%%
Zebrafish embryo & & & & \\ 
$t=625$ & 1 & 10 time points from $t=625$ to $t = 640$, $\sim~10,000$ cells each & $\alpha=60\mu m$ & Ref.~\cite{Keller2008} \\  
$t=775$ & 1 & 10 time points from $t=775$ to $t = 790$, $\sim~13,800$ cells each & \ditto & \ditto \\ 
$t=925$ & 1 & 10 time points from $t=925$ to $t = 940$, $\sim~15,100$ cells each & \ditto & \ditto \\ 
$t=1075$ & 1 & 10 time points from $t=1075$ to $t = 1090$, $\sim~14,800$ cells each & \ditto &
\ditto \\ 
$t=1225$ & 1 & 10 time points from $t=1125$ to $t = 1140$, $\sim~15,000$ cells each & \ditto & \ditto \\ 
$t=1375$ & 1 & 10 time points from $t=1375$ to $t = 1390$, $\sim~15,100$ cells each & \ditto &
\ditto \\ 
\hline& \\[-2ex]
%%%%%%%%%%%%%%%%%%%%%%%%%%%%%%%
\emph{D. melanogaster} embryo & 14 & evenly spaced time points from $t=270$ to $t=630$ m.p.f. Contains growing cell numbers from 11,593 to 27,026 & $\alpha=80\mu m$ & Ref.~\cite{Keller2010} \\ 
\hline& \\[-2ex]
%%%%%%%%%%%%%%%%%%%%%%%%%%%%%%%
\emph{P. mammillata} embryo  & 1 & 145 frames containing 250-700 cells across 5 experiments & $\alpha=80 \mu$m & Ref.~\cite{AscidianScience} \\ 
\hline& \\[-2ex]
%%%%%%%%%%%%%%%%%%%%%%%%%%%%%%%
\emph{C. elegans} embryo  & 1 & 6 experiments at $\sim$350 cell stage & $\alpha=8 \mu$m & Ref.~\cite{Cao2020} \\ 
\hline& \\[-2ex]
%%%%%%%%%%%%%%%%%%%%%%%%%%%%%%%
Human cancer organoid  & 1 & 16 experiments at $\sim$200 cell stage & $\alpha=50 \mu$m & Ref.~\cite{GuoOrganoid} \\ 
\hline& \\[-2ex]
%%%%%%%%%%%%%%%%%%%%%%%%%%%%%%%
Glassy dynamics  & 3 & 4096 & periodic & Ref.~\cite{GoogleGlassy} \\ 
\hline& \\[-2ex]
%%%%%%%%%%%%%%%%%%%%%%%%%%%%%%%
Open-cell polyurethane foam  & 1 & 18,000 pores& $\alpha = 300 \mu m$  & Ref.~\cite{OpenCellFoam} \\ 
\hline& \\[-2ex]
%%%%%%%%%%%%%%%%%%%%%%%%%%%%%%%
Fluid foam simulation & 1 & 21,000 bubbles  & $\alpha = 0.06$ a.u.  & Ref.~\cite{FoamSimulation} \\ 
\hline& \\[-2ex]
%%%%%%%%%%%%%%%%%%%%%%%%%%%%%%%
\emph{Arabidopsis thaliana} apical meristem  & 6 & 20 time points each containing $\sim2,000$ cells,  & $\alpha = 20\mu m$  & Ref.~\cite{Jonsson2016} \\ 
\hline& \\[-2ex]
%%%%%%%%%%%%%%%%%%%%%%%%%%%%%%%
Snowflake yeast  & 1 & 20 time points each containing $\sim100$ cells,  & $\alpha = 25\mu m$  & Ref.~\cite{Goldstein2022} \\ 
\hline& \\[-2ex]
%%%%%%%%%%%%%%%%%%%%%%%%%%%%%%%
Random packings & & & & \\ 
Sphere packing & 5 &  $10,000$  & periodic & Simulated using Ref.~\cite{Donev2005}\\ 
1:1:4 ellipsoid packing & 5 &  $10,000$  & periodic & ---\texttt{"}---\\ 
1:4:4 ellipsoid packing & 5 &  $10,000$  & periodic & ---\texttt{"}---\\ 
1:2:3 ellipsoid packing & 5 &  $10,000$  & periodic & ---\texttt{"}---\\ 
Polydisperse packing & 5 &  $10,000$  & periodic & ---\texttt{"}---\\ 
\hline& \\[-2ex]
%%%%%%%%%%%%%%%%%%%%%%%%%%%%%%%
Misc. & & & & \\ 
Poisson-Voronoi & 5 &  $10,000$  & periodic & This study\\  
Diffusion limited aggregation  & 5 & 10,000 & $\alpha=6.5$ a.u. & Simulated with Ref.~\cite{DLA_code} \\
Star survey data & 1 &  $110,000$  & $\alpha= 20$ parsecs & Refs~\cite{HYGDatabase,1997HIPPARCOS, 1995YALE, 1991GLIESE} 

\end{longtable}
\end{footnotesize}